\documentclass[a4paper,fleqn,usenatbib]{mnras}

\usepackage{mathptmx}
\usepackage{txfonts}  

\usepackage[T1]{fontenc}
\usepackage{ae,aecompl}

\usepackage{graphicx}	
\usepackage{xcolor}
\usepackage{caption}
\usepackage{threeparttablex}

\newcommand\etal{{~et\,al.\ }}

\newcommand\ngc{\ensuremath{N_{\rm GC}}}

\newcommand\mo{$M_{\odot}$}
\newcommand\sersic{S\'{e}rsic}
\newcommand\M{\ensuremath{\mathcal M}}

\newcommand\Mstar{\ensuremath{\mathcal M_{\star}}}

\newcommand\Mdyn{\ensuremath{\mathcal M_{dyn}}}
\newcommand\Mhalo{\ensuremath{\mathcal M_{Halo}}}
\newcommand\Mgc{\ensuremath{\langle{\mathcal M_{\rm GC}}\rangle}}
\newcommand{\Regc}{$R_{\rm e, GCS}$}
\newcommand{\ReGal}{$R_{\rm e, GAL}$}
\newcommand{\lamR}{$\lambda_{\rm R}$}

\title[Globular Cluster Systems of  Relic Galaxies]{Globular Cluster Systems of Relic Galaxies}

\author[Alamo-Mart\'inez et al.]{Karla A. Alamo-Mart\'inez,$^{1,2}$\thanks{E-mail: alrakomala@gmail.com}
Ana L. Chies-Santos,$^{1}$
Michael A. Beasley,$^{3}$
\newauthor
Rodrigo Flores-Freitas,$^{1}$
Cristina Furlanetto,$^{4}$
Marina Trevisan,$^{1}$
\newauthor
Allan Schnorr-M\"uller,$^{1}$
Ryan Leaman,$^{5}$
Charles J. Bonatto$^{1}$
\\
$^{1}$Departamento de Astronomia, Instituto de F\'isica, Universidade Federal do Rio Grande do Sul (UFRGS), Porto Alegre, R.S, Brazil\\
$^{2}$Departamento de Astronom\'ia, Universidad de Guanajuato, Apartado Postal 144, 36000, Guanajuato, Guanajuato, Mexico\\
$^{3}$Instituto de Astrof\'isica de Canarias, Calle V\'ia L\'actea, La Laguna, Spain\\
$^{4}$Departamento de Física, Instituto de F\'isica, Universidade Federal do Rio Grande do Sul (UFRGS), Porto Alegre, R.S, Brazil\\
$^{5}$Max-Planck Institut f\"ur Astronomie, K\"onigstuhl 17, D-69117 Heidelberg, Germany
}

\date{Accepted 2021 February 18. Received 2021 February 17; in original form 2020 December 8}

\pubyear{2019}

\begin{document}
\label{firstpage}
\pagerange{\pageref{firstpage}--\pageref{lastpage}}
\maketitle

\begin{abstract}
We analyse the globular cluster (GC) systems of a sample of 15 massive, compact early-type galaxies (ETGs), 13 of which have already been identified as good relic galaxy candidates 
on the basis of their compact morphologies, old stellar populations and stellar 
kinematics.
These relic galaxy candidates are likely the nearby counterparts of high redshift {\it red nugget} galaxies. 
Using F814W ($\approx$~$I$) and F160W ($\approx$~$H$) data from the WFC3 camara onboard the {\it Hubble Space Telescope} 
 we determine the total number, luminosity function, specific frequency, colour and spatial distribution of the GC systems. 
We find lower specific frequencies ($S_N<2.5$ with a median of $S_N=1$) than ETGs of comparable mass. This is  consistent with a scenario of rapid, early dissipative formation,  with relatively low levels of accretion of low-mass, high-$S_N$ satellites. 
The GC half-number radii are compact, but follow the relations found in normal ETGs.
We identify an anticorrelation between the specific angular momentum ($\lambda_R$) of the host galaxy  and the (I-H) colour distribution width of their GC systems. Assuming that $\lambda_R$ provides a measure of the degree of dissipation in massive ETGs, we suggest that the (I-H) colour distribution width can be used as a proxy for the degree of complexity of the accretion histories in these systems. 
\end{abstract}



\section{Introduction}

Massive early-type galaxies (ETGs) are host to some the most extreme formation events, and are the end products of complex processes such as gas accretion, in-situ star formation, hierarchical merging, tidal interactions, and secular evolution (\citealt{trager00}, \citealt{thomas05}, \citealt{barbera13}, \citealt{salvadorrusinol00}). ETGs experience  dramatic size evolution over cosmic history (e.g.\,\citealt{trujillo06}, \citealt{buitrago08}, \citealt{vandokkum10}), have the most extreme initial mass functions (i.e. weighted to low-mass stars, \citealt{vc10}, \citealt{barbera13}) and host the most massive black-holes in the Universe (\citealt{FM00}, \citealt{mcma13}).
The rich variety of mechanisms that shape ETGs together with the fact that these galaxies contain $\sim$70\% of the total stellar budget in the local Universe (\citealt{Fukugita1998}), turn these galaxies into attractive laboratories to test galaxy formation theories.  

There is theoretical and observational evidence that massive ETGs ($M_* > 10^{11}$~M$_\odot$) in the local Universe have gone through two major formation phases (eg. \citealt{hopkins2009}, \citealt{oser10}, \citealt{hill17}). In the first phase, an {\it in situ} component is formed through dissipational processes, creating a massive and compact central component. This first phase happens rapidly (< 1 Gyr)  and early (at $z\gtrsim2$) (\citealt{zolotov2015}), and gives rise to a population of passively evolving objects seen at high redshift (\citealt{daddi05}, \citealt{schreiber18}, \citealt{valentino20}), sometimes termed 
``red nuggets" (\citealt{damjanov09}). A second phase, dominated by minor mergers,  then builds-up predominantly  the outer regions  of ETGs (\citealt{Naab2009}, \citealt{Johansson2009}). This process grows massive galaxies in size and mass, and presumably lowers their central densities (\citealt{hilz12}, \citealt{hilz13}). Simulations suggest that the second phase may initially begin concurrently with the first, but continues to operate until the present day (e.g., \citealt{wellons15}, \citealt{furlong17}.
For the purposes of this paper, we refer to an  "accretion event" as any merger that has occurred at $z < 2$.

A consequence of this formation pathway is that  a small fraction of these red nuggets are expected to survive ``frozen" until the present-day Universe (\citealt{trujillo09}, \citealt{poggianti13}, \citealt{spiniello20}) . These objects are thus considered relics of the first phase of ETG formation. In fact, such relic candidates have been reported to exist relatively nearby (\citealt{y17}, \citealt{FM17,FM18}).
As relic galaxies, they are expected to  have uniformly old stellar populations ($>10$\,Gyr, \citealt{y17}), and be extremely compact, as is the case for the prototypical relic galaxy  NGC\,1277 (\citealt{trujillo14}) and also Mrk\,1216 and PGC\,032873 (\citealt{FM17}). In addition, stellar population studies indicate that relics such as  NGC\,1277 have extremely bottom-heavy stellar initial mass functions (IMF) which do not vary with radius. This is consistent with the picture that this class of objects generally go on to form the central regions of massive ETGs (\citealt{mn15b}).
Having accreted very few satelites, one might expect that relic galaxies would be found preferentially in relative isolation. However, a handful of relic candidates (\citealt{y17}, ) are found in the Perseus cluster. This result finds support in the work of \cite{poggianti13} and \cite{Peralta2016} who find massive compact galaxies are preferentially found in central regions of the most dense environments. This may suggest that the conditions in early proto clusters may have been especially favourable to preserve these systems. In the particular case of NGC~1277 in Perseus, it is possible that its proximity to NGC~1275 (Perseus A), has allowed NGC 1277 to evolve with little  accretion. NGC 1275 may have acted (and continues to act) as a main attractor for any circumgalactic material.

Identification and characterization of relic galaxies is crucial to test the two-phase model, and offers the opportunity to study red nugget galaxies in the nearby Universe. 
In addition to the chemo-dynamical study of galaxy stellar populations, globular clusters (GCs) have proven to be uniquely powerful tools to trace the structure and history of galaxies (\citealt{brodie06}, \citealp{pfeffer18}, \citealp{beasley20}). 

GC systems are generally thought to be old (most of them with ages $>10$\,Gyr, \citealt{strader05}, \citealt{chies11}), and contain the imprints of  the initial conditions of galaxy formation.  
GC systems of ETGs in the local Universe generally show complex colour distributions. 
Because GCs are generally old and co-eval, differences in color are attributed to differences in metallicity, such that  metal-poor GCs have ``blue" colours and  metal-rich GCs are ``red". This correspondence has been confirmed with spectroscopy (e.g., \citealt{beasley08}, \citealt{Caldwell2011}, \citealt{usher12}).
Furthermore, the fraction of red GCs decreases as we move to lower galaxy luminosities, reaching the extreme faint of dwarfs that have predominantly blue GCs  (e.g. \citealt{peng06}). 
This is explained by invoking the mass-metallicity relation where the metal-poor GCs are formed in low mass galaxies, while the metal-rich GCs are formed in more massive proto-galaxies (\citealt{cote98}; \citealt{strader05}). In the two-phase galaxy formation context, the red metal-rich GCs are thought formed predominantly {\it in-situ}, while an important fraction of  blue, metal-poor GCs are  accreted from dwarf galaxies. 

Recently, \citet{beasley18} found that NGC\,1277 contains almost exclusively red GCs suggesting that the galaxy has largely bypassed the second-phase of massive ETG formation, where  ETG assembly should be dominated by minor merging events which would have brought in blue metal-poor GCs. The GC results for NGC\,1277 suggest an accreted mass fraction of $\sim12\%$ in this galaxy. By contrast, massive ETGs are generally estimated to have accreted somewhere between 50--90\% of their stars, with the rest comprising of an {\it in-situ} component (\citealt{oser10}; \citealt{Navarro-Gonzalez2013}). 

Here, we present the first systematic study of GC populations in relic candidate galaxies. 
This paper is structured as follows. In Section 2 we introduce the data, photometric analysis and GC candidates selection. In Section 3 we analyse the spatial distribution, luminosity function and total number of the GC systems. 
In Section 4 we look for correlations between the determined GC system properties and host galaxy properties from the literature, such as, stellar mass, specific angular momentum, stellar velocity dispersion and metallicity. In Section 5 we discuss and summarise the results.

\section{Data}

In this work we analyze a sample of 15  massive, compact early-type galaxies identified in the Hobby-Eberly telescope Massive Galaxy Survey (HetMGS; \citealt{vdBosch15}). Table\,\ref{tab:Table1_Gals} contains basic properties of our sample galaxies. These galaxies were initially selected in order to resolve their black hole ``sphere of influence", $r_i\equiv~GM \sigma^{-2}$, where $M$ is the black hole mass and $\sigma$ is host galaxy velocity dispersion. Such a selection yields a strong bias in favour of massive, compact and high velocity dispersion systems. Of the 15 HetMGS galaxies considered here, 13 have been identified as good relic galaxy candidates on the basis of their compact  morphologies, old stellar populations and stellar kinematics (\citealt{y17}; \citealt{FM17}). Their properties are very similar, though not as extreme, as the relic galaxy archetype NGC~1277 (\citealt{trujillo14}, \citealt{mn15b}, \citealt{beasley18} ).

We use archival {\it Hubble Space Telescope} ({\sl HST}) data from the programme GO: 13050 (PI: van den Bosch). The observations were taken with the WFC3 camera with the detectors UVIS (FOV 162$\times$162 arcsec) and IR (FOV 123$\times$136 arcsec) through filters F814W ($\approx$~$I$) and F160W ($\approx$~$H$), respectively. 
The data is homogeneous and all the targets have total integration time of 500\,s for $I$ and 1400\,s for $H$. 
We downloaded the data from The Hubble Legacy Archive\footnote{https://hla.stsci.edu/} which offers reduced data from {\sl HST} following the standard procedure.

The FWHM for $I$ and $H$ are 0.1 and 0.22 arcsec, respectively. We use the AB photometric system, with zero-points in $I$ and $H$, $zp_I$=25.12 and $zp_H$=25.95 mag, respectively. The Galactic extinction toward each galaxy in both bands are from \citet{Schlafly_Finkbeiner2011}. 
For consistency with \citet{y17}, we adopt a cosmology with a Hubble constant of $H_0=70.5\,km\,s^{-1}\,Mpc^{-1}$, a matter density of $\Omega_M=0.27$, and dark energy density of $\Omega_{\Lambda}=0.73$.

\subsection{Photometry}

In order to detect sources close to the centre of the galaxies, first we need to subtract the galaxy light. This is because the steepness of the galaxy luminosity is reflected as drastic changes in the background, hampering the detection of sources. To create the galaxy model we use the tasks {\sc ellipse} and {\sc bmodel} from IRAF (\citealt{Tody1986}) and fit elliptical isophotes to the surface brightness of the galaxies. 
For such procedure, we mask all the sources detected (except the galaxy to be fitted) using the SEGMENTATION MAP from a first SExtractor (\citealt{Bertin_Arnouts1996}) run. 
We then create models for both $I$ and $H$ bands and subtract them from the original images. 

When inspecting the subtracted images, we notice significant residuals in the central regions of most of the sample galaxies, this is mainly because of dust (disks or lanes) or complex structures like rings. Another manuscript is being drafted with an in-depth structural analysis of this sample of galaxies (Flores-Freitas, \textit{in prep}).

We perform source extraction on the galaxy subtracted image, by running SExtractor independently for each band with a set of parameters optimized for point source detection. We use a background grid size BACK\_SIZE=32\,pix and FILTERSIZE=3, and an effective $\sigma$ detection larger than 5 for both bands.  
Then, we apply a first cut to select objects with good quality in photometry by rejecting objects with saturated pixels, truncated by being close to the edges of the image, or, with magnitude error larger than 0.2 mag. Moreover, we estimate the concentration parameter $C_{4-10}$, which is the difference in magnitude within an aperture of 4 and 10 pixels in diameter, as it seems to be a better indicator of the luminosity profile for the faintest objects in comparison to the SExtractor CLASS\_STAR parameter (\citealt{Peng2011}, \citealt{Cho16}). 

\begin{table*}
\begin{threeparttable}
	\centering
	\caption{Galaxy sample parameters:  (1) Galaxy name. (2) Distance. (3) Environment estimate - galaxy number density ($\rho$) within the volume containing the two closest neighbours. (4) Environment according to NED. (5) Integrated Magnitude in the V-band. (6) Reddening. (7,8) Aperture corrections in the I and H bands.}
	\label{tab:Table1_Gals}
	\begin{tabular}{lrrrrrrr}
\hline
 Galaxy    &   D [Mpc] & $\rho$ [\#/Mpc$^3$] & E & $M_{V,GAL}$ &    $A_V$ &   ap\_c I &   ap\_c H \\
 (1) & (2) &  (3) & (4) &  (5) &  (6) &  (7) & (8) \\
\hline
 NGC 0384  &             59 &  0.08   & pair/group &   -21.17 &  0.17 &      -0.18 &      -0.27 \\
 NGC 0472  &             74 &  3.17  & - &   -21.61 &  0.13 &      -0.15 &      -0.20 \\
 MRK 1216  &             94 & 0.11  & - &    -22.12 &  0.09 &      -0.18 &      -0.19 \\
 NGC 1270  &             69 &  7.76 & group/cluster Abell\,426*  &    -21.85 &  0.45 &      -0.16 &      -0.19 \\
 NGC 1271  &             80 & 7.32  & group/cluster Abell\,426 &    -22.04 &  0.45 &      -0.16 &      -0.18 \\
 NGC 1281  &             60 &  1.64 &  group/cluster Abell\,426 &  -21.26 &  0.46 &      -0.17 &      -0.17 \\
 NGC 1282  &             31 &  0.35 & group/cluster Abell\,426 &     -20.32 &  0.46 &      -0.18 &      -0.20 \\
 UGC 2698  &             89 & 0.83  & - &   -22.36 &  0.40 &      -0.17 &      -0.18 \\
 NGC 2767  &             74 & 0.96  & group  &   -21.06 &  0.05 &      -0.20 &      -0.21 \\
 UGC 3816  &             51 &  0.27 & group &    -21.66 &  0.17 &      -0.33 &      -0.24 \\
 NGC 3990  &             15 &  1.37 & pair/group &    -18.95 &  0.05 &      -0.19 &      -0.25 \\
 PGC 11179 &             94 & 4.86  & group/cluster Abell\,400 &    -21.70 &  0.51 &      -0.17 &      -0.24 \\
 PGC 12562 &             67 &  1.94 & group/cluster Abell\,426 &     -20.95 &  0.46 &      -0.18 &      -0.30 \\
 PGC 32873 &            112 & 0.43  & pair &     -22.12 &  0.04 &      -0.31 &      -0.23 \\
 PGC 70520 &             72 &  0.01 & - &      -21.75 &  0.26 &      -0.16 &      -0.19 \\ 
\hline
\end{tabular}
\begin{tablenotes}
   \item[*] Perseus cluster  
  \end{tablenotes}
\end{threeparttable}
\end{table*}
 
\subsection{Globular Cluster Candidates Selection}
\label{sect:sourceDect_and_completeness}

We use SExtractor output MAG\_AUTO as the actual magnitude of the objects, but to estimate the color we use aperture photometry within a radius of 4 pixels (0.16$^{\prime\prime}$) and apply the respective aperture correction. 
The aperture correction is estimated by constructing the curve of growth for multiple point sources, measuring the magnitude within different apertures (radius of 2,4,5,8,11,14,16 pixels) until it reaches a constant value, representing the total magnitude. Since our target galaxies are located at different distances, a fixed aperture corresponds to a different physical size at the source distance. In Table\,\ref{tab:Table1_Gals} we present the aperture corrections for each galaxy in the I and H bands. 
Finally, the sample of point sources that are taken as GC candidates are the ones detected independently in both bands (matching the coordinates within a radius $<1$\,arcsec).
Also, the GC candidates must have magnitudes according to the expected luminosity function of GCs (see Section\,\ref{sect:GCLF} for more details), and color -0.5 < $I-H$ < 1.5 which is the expected color for GCs (\citealt{Cho16}).

In order to have an accurate estimation of the GCs magnitude distribution and the total number of GCs for each galaxy, we need to determine the detection completeness curve as function of magnitude. 
To estimate the completeness curve we construct the point-spread function (PSF) using the {\sc Tiny Tim} \footnote{http://tinytim.stsci.edu/cgi-bin/tinytimweb.cgi} web interface.  We add 20,000 artificial stars with a range in magnitude between 22 and 27 in both bands, and color $I-H$ = 0. We mimic the source detection as in the original images by applying the same source detection criteria in each band independently, and then select the ones that match in coordinates. 
We proceed by fitting a Pritchet function (\citealt{Fleming1995}) to the ratio of recovered to created artificial stars for the $I$ band, as it is the band with reliable parameters available for GC populations (\citealt{Peng2011}, \citealt{Cho16}). 
As the observing strategy is the same for all the galaxies, the recovered parameters are similar but not exactly the same as the detection limit is affected by the brightness of the galaxy.  The magnitude in the $I$-band where 50\% of the artificial stars are recovered is in the range of 25.1 and 25.5, and with slopes in the range 3-5 for all the sample.



\section{Analysis}
\subsection{Spatial Distribution of Globular Cluster Candidates}
\label{sect:GCS_sersic}

In order to estimate the amount of contaminants by Milky Way stars and background galaxies, and to quantify the spatial incompleteness, we parametrise the spatial distribution of the GC candidates by determining the surface number density of the GC candidates as function of radius. 
We measure the number of GC candidates inside circular concentric annuli with a constant width of 16\,arcsec each, and normalise it by the effective area. We then fit a single \sersic\ function (\citealt{Sersic1968}) including a background component: 

\begin{equation}
\Sigma_{GC}(r)= \Sigma_0\exp\left\{-b_n \left[ {\left( \frac{r}{{R_{\rm e, GCS}}}\right) }^{1/n_{\rm GCS}}- 1\right] \right\}  + BG_{ps}
\label{eq:GCS_sersic}
\end{equation}

\begin{figure}
\centering
\includegraphics[width=0.5\textwidth]{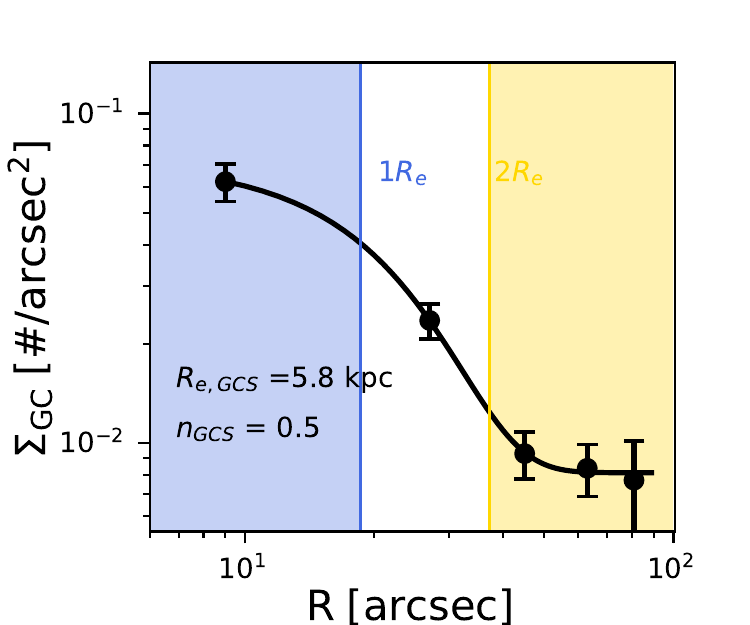}
\caption{Surface number density of the GC candidates as function of radius for the galaxy NGC\,1270. The error bars represent the poissonian error for the number of counts and weighted by the  effective area of each annulus. The solid line shows the best \sersic\ fit. The blue shaded region shows the {\it inner} population which contains mostly GCs with high certainty, while the yellow shaded region shows the {\it outer} population which might be strongly affected by contaminants. }
\label{fig:NGC1270_GCspatialDist}
\end{figure}

\noindent
where \Regc\ is the radius that contains half of the GC population; $\Sigma_0$ is the surface number density at $R_{\rm e, GCS}$; $n_{\rm GCS}$ is the S\'ersic index; and $BG_{ps}$ is the surface number density of background point sources. 
The fit is performed leaving all the parameters free but constraining $n_{\rm GCS}$ to be between 0.5 and 4, as are the expected values for  disky and spheroidal galaxies (\citealt{Kormendy2009}), and \Regc\ to be larger than the effective radius of the galaxy ($R_{\rm e, GAL}$) as the physical extension of the GC system is generally more extended than that of the starlight (\citealt{Rhode2007}; \citealt{Kartha14}). 
The error bars are estimated assuming a poissonian error for the number of counts and weighted by the  effective area. 
As an example, in Figure\,\ref{fig:NGC1270_GCspatialDist} we show the GC surface number density for the galaxy NGC\,1270, and its best \sersic\ fit. 
For all  the galaxies we obtained \Regc\ between 2 and 3 times \ReGal, in agreement with the literature (\citealt{Kartha14}; \citealt{Forbes2017}; \citealt{Hudson_Robison2018}).  
In Table\,\ref{tab:Table2_GCS} we present the measured \Regc\ values. 

From inspection of the Sersic fits, we define two subsamples: an {\it inner} population (sources within 1\Regc, blue shaded region in Fig.\,\ref{fig:NGC1270_GCspatialDist}), whose sources with high certainty are dominated  by GCs, and an {\it outer} population (sources out of 2\Regc, yellow shaded region in Fig.\,\ref{fig:NGC1270_GCspatialDist}), which might be heavily contaminated by  Milky Way stars and background galaxies. These subsamples are defined so that the inner subsample is well above the  level of background contaminants.

\subsection{Luminosity Function and total GC number}
\label{sect:GCLF}

\begin{figure}
\centering
\includegraphics[width=0.5\textwidth]{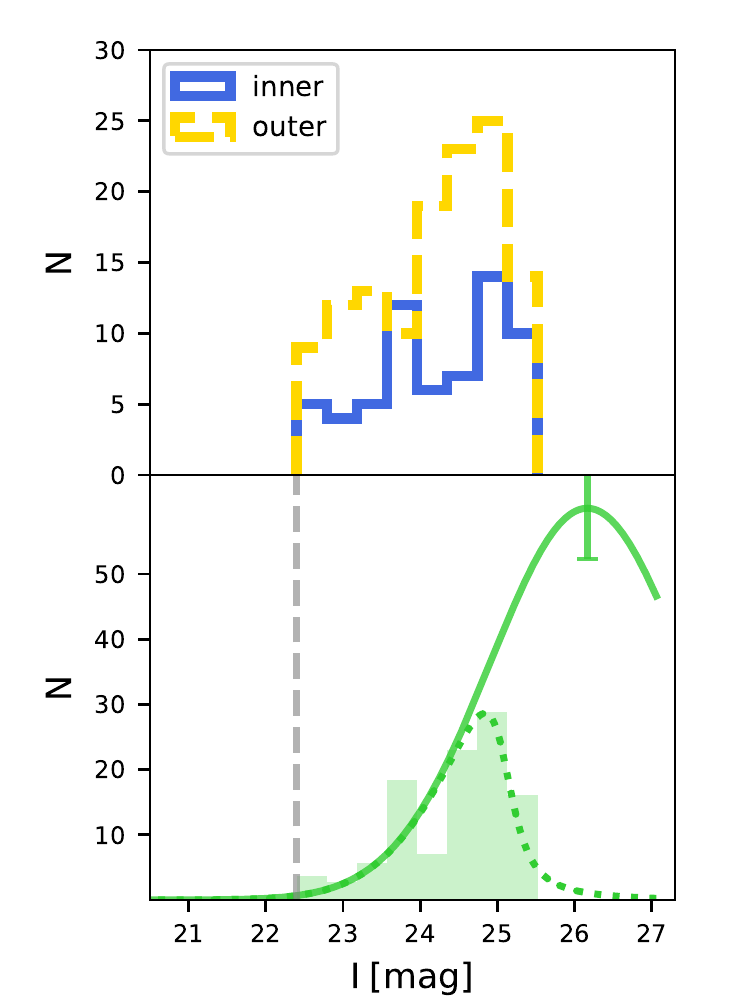}
\caption{GC luminosity functions for galaxy NGC\,1270. Top panel: Luminosity function for the {\it inner} and {\it outer} subpopulations. Bottom panel: Luminosity function of the {\it inner} population minus the luminosity function of contaminants within 1\Regc (corrected by area), multiplied by two, as within 1${R_e}$ we have half of the total population. 
The dotted line shows the best Gaussian*Pritchet function, the solid line is the corresponding single Gaussian, and, the gray dashed vertical line indicates the bright cut for selected sources (${M_I}^0$-3${\sigma_I}^{\rm GCLF}$). 
}
\label{fig:GCLF_NGC1270}
\end{figure}

The globular cluster luminosity function (GCLF) is a property that has been extensively studied among galaxies with different morphological type, mass and environment, being well described by a Gaussian (\citealt{Harris1991}), albeit some departures form gaussianity in the faint end (\citealt{Jordan2007}), not relevant for this study. 
The peak or also called turnover (${M}^0$) and width (${\sigma}^{\rm GCLF}$) of the GCLF are known to depend on the luminosity of the galaxy, where brighter turnover and broader distributions are observed in brighter galaxies (\citealt{Kundu_Whitmore2001}; \citealt{Jordan2006}). For ETGs the values are ${M_V}^0\approx-7.4$ and  ${M_I}^0\approx-8.5$ in Johnson-Cousins photometric system. On the other hand, although ${\sigma}^{\rm GCLF}$ varies with the luminosity of the galaxy, it does not vary among bands (\citealt{Kundu_Whitmore2001}) where for the most massive galaxies is ${\sigma_V}^{\rm GCLF}\approx1.4$. 

We adopt a GCLF turnover and width (AB system) dependent on the galaxy luminosity $M_{V,GAL}$ with the form: 

\begin{equation}
{M_I}^0=-7.4 + 0.04(M_{V,GAL} + 21.3) -1.04 + 0.436 
\end{equation}

\begin{equation}
{\sigma_I}^{\rm GCLF}=1.2 - 0.1(M_{V,GAL}+21.3)
\end{equation}

\noindent
following (\citealt{jordan06}, \citealt{villegas2010} and \citealt{Harris13}) on the dependency of ${M_V}^0$ as function of the galaxy luminosity, and assuming ${M_V}^0 - {M_I}^0=1.04$ which is a mean value measured for 28 ETGs according to \citealt{Kundu_Whitmore2001}, and finally to convert from Johnson-Cousins to AB photometric system we add 0.436 (\citealt{Sirianni2005}). 

Using this transformation, for a massive galaxy with $M_{V,GAL}\sim-24$ (as it is for M87) we recover ${M_I}^0=-8.112$ in agreement with \cite{jordan07}. On the other hand, for dwarf galaxies with $M_{V,GAL}\sim-18$, we recover ${M_I}^0=-7.872$ in agreement with \citet{Miller_Lotz2007}.

As mentioned in Section\,\ref{sect:sourceDect_and_completeness}, we selected point sources fainter that 3${\sigma_I}^{\rm GCLF}$ from ${M_I}^0$, together with a selection by color -0.5 < $I-H$ < 1.5 (\citealt{Cho16}), in order to reduce the contamination by Milky Way stars and background galaxies. 

To derive the GCLF of our studied galaxies, we estimate the luminosity function for the {\it inner} and {\it outer} subpopulations. 
This ultimately allows us to create an {\it inner} GCLF clean from contaminants. The luminosity function of the outer region (dominated by contaminants) is normalized to the area of the inner region, representing the luminosity function of contaminants.
Then, our final GCLF for each galaxy is the luminosity function of the {\it inner} population minus the luminosity function of contaminants within 1\Regc, multiplied by two, as within 1${R_e}$ we have half of the total population.

As the GCLF that we recover is the convolution of the intrinsic GCLF and the incompleteness Pritchet function (determined in Sect\,\ref{sect:sourceDect_and_completeness}), in order to obtain the total number of GCs, we estimate the area under the expected Gaussian GCLF by fitting the product of a Gaussian and the Pritchet function, with the expected ${\sigma_I}^{\rm GCLF}$ and ${M_I}^0$. 
In Figure\,\ref{fig:GCLF_NGC1270} we show the luminosity functions for the {\it inner} and {\it outer} subpopulations (top panel), as well as the GCLF (bottom panel) in the $I$-band. We also show the best Gaussian*Pritchet function (dotted curve), and the corresponding single Gaussian (solid curve) for the galaxy NGC\,1270. 

Although the data is relatively deep, it is not enough to sample the whole range of the GCLF, the faint limit we reach is about $\sim$1 mag brighter than ${M_I}^0$ (depending on the distance to the galaxy). In Table\,\ref{tab:Table2_GCS} we show the total number of GCs corrected by incompleteness in area and photometry, and the corresponding error propagating the uncertainty in the fitted Gaussian amplitude.  \\

\section{Results}

In this section we explore different correlations between the GC System (GCS) and host galaxy parameters derived in this work and from the literature. 

\subsection{GCs Color ($I - H$) correlations}
\label{sect:GC_colors}

\begin{figure}
\centering
\includegraphics[width=0.5\textwidth]{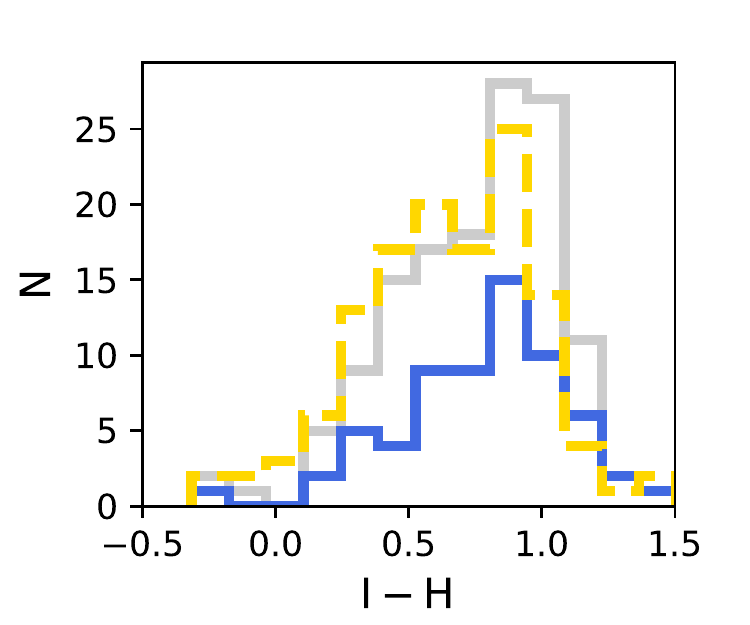}
\caption{GC candidates color $I-H$ distributions for the population within 1\Regc\ (solid blue), outside 2\Regc\ (dashed yellow), and from 0 to 2\Regc\ (solid gray), for the galaxy NGC\,1270. }
\label{fig:colorHist_NGC1270}
\end{figure}

The colours of the GC populations contain significant amounts of embedded information. They give us insights into ages and metallicities of the GCs (\citealt{brodie06}), key parameters to infer the star-formation history of the host galaxy. Spectroscopic analyses indicate that most GCs are old and co-eval (\citealt{cohen03}; \citealt{Puzia2005}; \citealt{beasley08}; \citealt{Caldwell2011}), implying that color differences are mainly due to differences in metallicity. 
Unlike the relative  homogeneity of the GCLF shape, the color distribution of GCs shows more variety among different galaxies. In optical colors, more massive galaxies exhibit more complex GC color distributions (dominated by bimodal distributions with different blue-to-red fractions, see \citealt*{Lee_Yoon2019}) and turn gradually unimodal with a blue dominant population as the mass of the galaxy decreases (\citealt{peng06}).

\begin{figure*}
\centering
\includegraphics[width=0.62\textwidth]{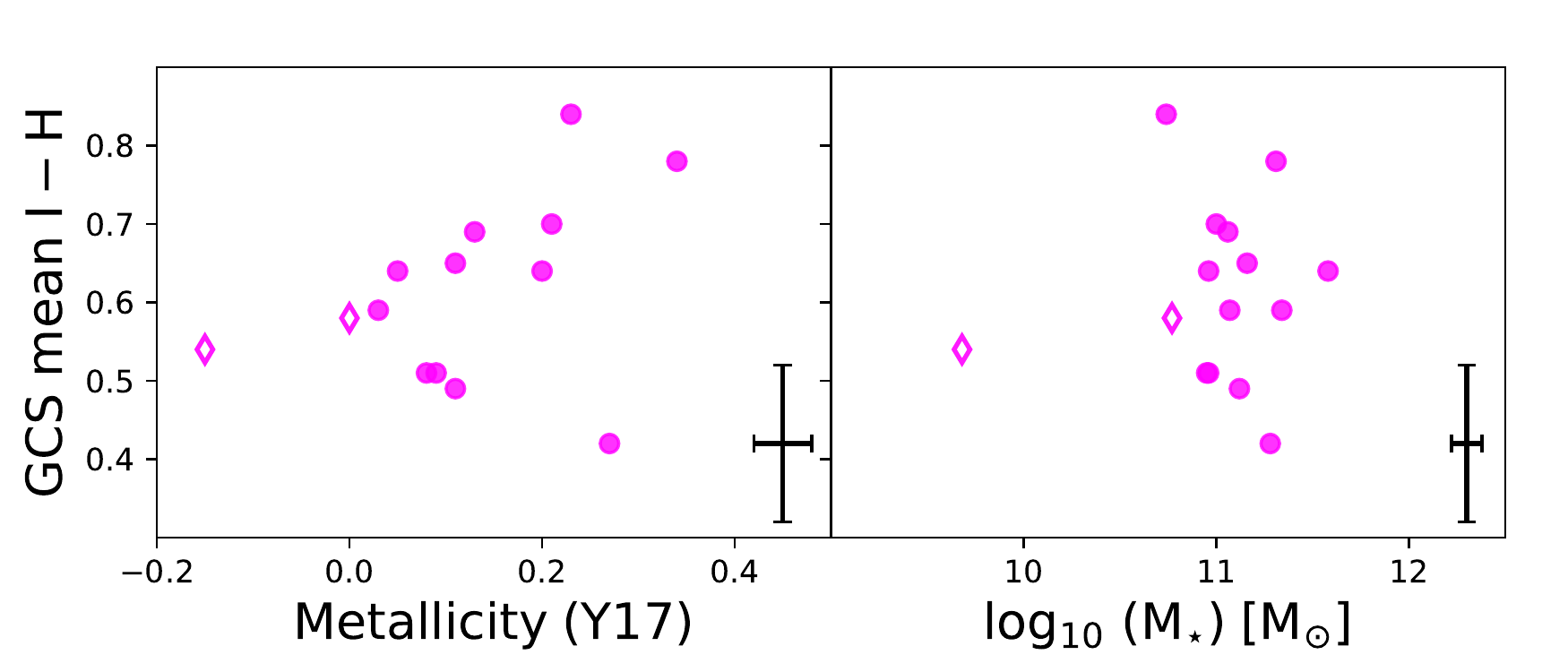}
\includegraphics[width=0.36\textwidth]{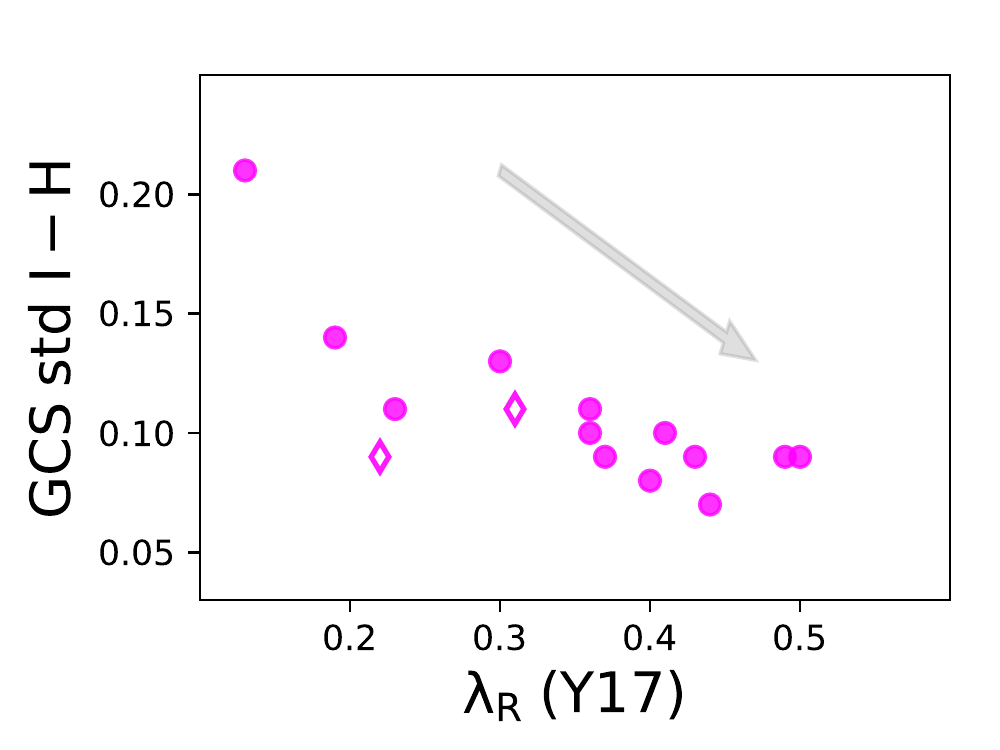}
\caption{Mean and standard deviation of $I - H$ distributions within interquartiles 1 and 3 for all the GC candidate sources (i.e. without background contamination subtraction) within 2\Regc.  Metallicity, \Mstar\ and \lamR\ are from \citet{y17}. 
The bottom right shows the typical error bars. The dots are the galaxies classified as relics and the diamonds as non-relics by \citet{y17}. The arrow indicates the direction at which the ``degree of relicness" of a given GC system increases (see text for details).}
\label{fig:meancolorGCs}
\end{figure*}

Even though the optical colors of GC systems have been extensively studied, their interpretation is still debated. 
Here, we do not attempt to study in detail the nature or the existence of bimodality of the near-IR color distribution of GCs (for that, see \citealt{blakeslee12}, \citealt{CS12}, \citealt{Cho16}) mainly because we are not sampling the entire population of GCs in any of the bands. Instead, we  perform a rough characterization of the $I-H$ color distribution of our sample by estimating the mean, the standard deviation, and the skewness of the GC candidates for each galaxy. The rationale here is that even such simple metrics can encode useful information about the GC systems and their host galaxies.
In order to avoid outliers, we calculate the mean and standard deviation within interquartiles 1 and 3 for all the GC candidate sources (i.e. without background contamination subtraction), for the {\it inner} (within 1\Regc), {\it outer} {outside 2\Regc, and from 0 to 2\Regc. } 

It is well known that the mean metallicity of a galaxy and its GCs increases for more massive galaxies  (\citealt{Kundu_Whitmore2001}; \citealt{usher12} -- and see the compilation in \citealt{beasley19}); together with the fact that the fraction of red GCs increases for brighter galaxies (\citealt{peng06}), we expect redder GCS colors for brighter galaxies. However, for a single galaxy, inner GCs have higher metallicities than the outer ones (\citealt*{Geisler1996}; \citealt{Harris2009}), thus, in order to avoid a bias for selecting only the inner GCs,  the GC color of our sample is determined using the population from 0 to 2\Regc\ where we have $\sim$90\% of the detected GC candidates. 
In Figure\,\ref{fig:colorHist_NGC1270} we show the color $I-H$ distribution for galaxy NGC\,1270, plotting the {\it inner}, {\it outer} and from 0 to 2\Regc\ subpopulations.

The skewness values reflect moderately skewed $I-H$ color distributions for five galaxies, while for the rest of the sample galaxies, the color distributions seem symmetrical (see Table\,\ref{tab:Table2_GCS}).
In Figure\,\ref{fig:meancolorGCs} we plot the mean $I-H$ GCS color versus galaxy metallicity and galaxy stellar mass (\Mstar) obtained from \citet{y17} where we can see a trend that more massive and metal-rich galaxies have GCs with on average redder mean $I-H$ colours, mimicking the effect of the GC peak metallicity, galaxy luminosity relation (see eg. \citealt{brodie06}). 
We also explore any dependence of the standard deviation of the $I-H$ GC color with galaxy properties, identifying an interesting possible correlation when plotting it versus the stellar angular momentum, \lamR, from \citet{y17}. We can see tantalizing evidence for an anti-correlation between \lamR\ and $I-H$ color standard deviation. This trend may be explained in a hierarchical scenario where successive merging events decrease  \lamR\ (\citealt{Rodriguez-Gomez2017}) while the final merged system in turn will tend to have a more complex, wider GC color distribution.  In this sense, the right hand panel of Figure\,\ref{fig:meancolorGCs} may show a progression of the ``degree of relicness" of a system, where the least evolved (most relic-like) systems will be located to the right of the plot with higher \lamR~ and a narrower colour width.

\subsection{GC total number correlations}

\begin{figure}
\centering
\includegraphics[width=0.4\textwidth]{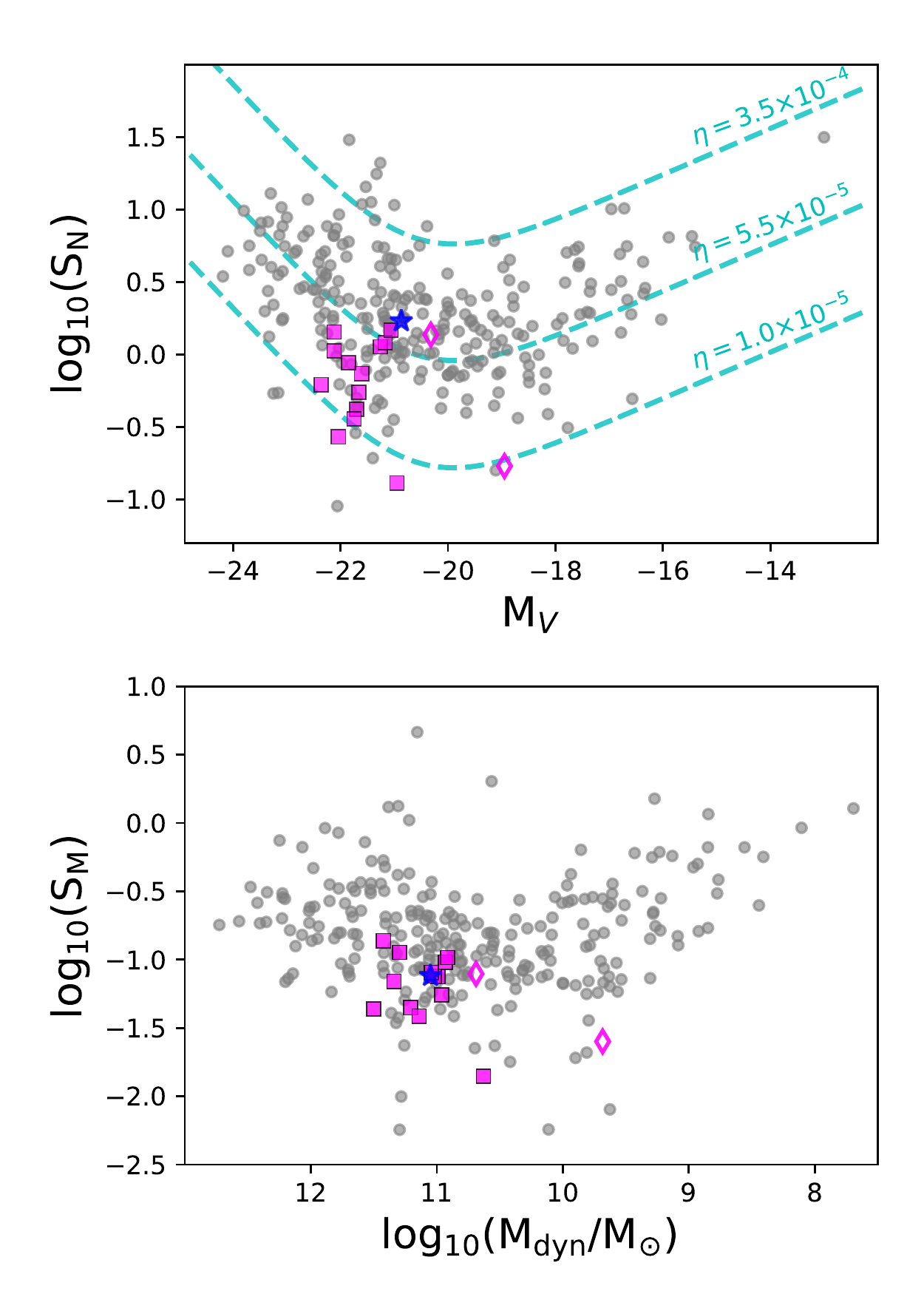}
\caption{Upper panel: The specific frequency $S_N$ as function of $M_V$ with magenta squares representing our sample galaxies classified as relics and the diamonds as non-relics by \citet{y17}, the gray dots are galaxies from literature (\citealt{Harris13}), and the blue star is the galaxy NGC\,1277 from \citet{beasley18}. The dashed cyan lines represent predictions from \citet{georgiev10} assuming fixed GC formation efficiency $\eta$ per total halo mass \Mhalo. Bottom panel: The specific mass $S_M$ versus dynamical mass as derived from Eq.~\ref{eq:Mdyn} (see text).}.
\label{fig:Sn_Sm}
\end{figure}

One of the first parameters defined to quantify the richness of a GC population is the called {\it specific frequency}, ${S_N}$ (\citealt{Harris_vdB1981}), which is the number of GCs per unit galaxy luminosity, normalised to a galaxy with absolute $V$ magnitude of -15:

\begin{equation}
S_N=10^{0.4(M_{V,GAL}+15)} .
\label{eq:Sn}
\end{equation}

\noindent
\citet{Harris_vdB1981} found values in the range 2$<S_N<$10 for elliptical galaxies, noticing that the number of GCs do not scale with the luminosity of the galaxy. Since then, the $S_N$ has been extensively studied among galaxies with different masses, morphological types and environments (\citealt{Harris01}; \citealt{peng08}; \citealt{georgiev10}, \citealt{Harris13} \citealt{AM17}), where the behaviour of  $S_N$ as function of $M_{V,GAL}$ can be described by an U-shape, where $S_N$ increases with luminosity on the bright side, and increases as the luminosity decreases on the faint side with a minimum of $S_N\sim1$ for galaxies around $L_V^\star$, and large scatter on the edges.

The fact that the number of GCs seems to scale linearly with the halo mass of the host galaxy (\citealt{Blakeslee1999}; \citealt{Kravtsov_Gnedin2005}; \citealt{Spitler_Forbes2009}; \citealt*{Hudson2014}; \citealt*{Harris2017}; \citealt{boylankolchin17};  \citealt{choksi19};  \citealt{elbadry19}) suggests that $S_N$ scales inversely with the stellar-to-halo mass relation, or the total star-formation efficiency, where high star-formation efficiencies would imply low values of $S_N$.  
On the other hand, because $S_N$ can be high in both giants and dwarf galaxies, but is uniformly low at the intermediate-mass regime, an explanation for the high $S_N$ values in giant galaxies is that the large amounts of GCs are an accreted population that once belonged to dwarf galaxies. Recently, \citet{beasley18} reported that the GCS of NGC\,1277, in the Perseus cluster is made up of almost exclusively red GCs, with only a small (<10\%) fraction of blue GCs by studying the optical color $g-z$ with deep HST imaging. This finding supports the idea from other studies based on the age of its stellar population (\citealt{FM18}) and its IMF (\citealt{mn15b}) that this galaxy is a relic. 
Furthermore, the $S_N$ of NGC\,1277 is low (<~2), in accordance with a high star-formation efficiency as would be for a rapidly early dissipative formation and a lack of a accretion of high-$S_N$, low-mass satellites.

In Figure\,\ref{fig:Sn_Sm} (upper panel) we show $S_N$ vs. $M_{V,GAL}$ for our sample and the compilation of  galaxies from \citet{Harris13}, which comprises a sample of 422 ETGs, S0s and late-type galaxies with $-24 < M_v < -14$. We find low values of $S_N$ for our sample, $< 2.5$ with a median of 1. 
Assuming that the behaviour of the bright side of $S_N$ vs. $M_{V,GAL}$ plot is dominated by accretion, and that the faint side is consequence of dissipative processes, then, relic galaxies would be the largest {\it seeds} formed by dissipative processes. 
A massive galaxy with low $S_N$ could also be the consequence of a post-starburst due to a recent gas accretion with high luminosity dominated by young stellar population, as it seems to be the case for the second brightest galaxy in Fornax cluster (\citealt{Liu2019}).

A quantity harder to derive than $S_N$ but with more physical meaning is the specific mass, $S_M$ (\citealt{peng08}), which is the fraction of baryonic mass turned into GCs (\M$_{GCS}$):

\begin{equation}
S_M=100\frac{\M_{GCS}}{\M_{baryonic}}
\label{eq:Sm}
\end{equation}

\noindent
where for non-star forming galaxies we can assume \M$_{gas}\sim0$ and then \M$_{baryonic}\sim$\Mstar, valid for our sample of early-type galaxies. 
As we know the total number of GCs, we just have to multiply by the mean GC mass, \Mgc. As mentioned in Section\,\ref{sect:GCLF}, the mean and standard deviation of the GCLF depend on the luminosity of the galaxy, being brighter and broader as the galaxy luminosity increases. This is reflected in larger \Mgc\ for more luminous galaxies. 
By using the relation ${M_V}^0=-7.4+0.04(M_{V,GAL}+21.3)$ and $(M/L)_V=2$ from \citet{Harris13}, we derive \Mgc. 
To estimate the dynamical mass, we use: 

\begin{equation}
\Mdyn=\frac{4R_{\rm e, GAL}\ {\sigma_{\rm e}^2}}{G}
\label{eq:Mdyn}
\end{equation}

\noindent
where \ReGal\ is the effective radius of the galaxy, and $\sigma_{\rm e}$ is the stellar velocity dispersion within \ReGal, both obtained from \citet{y17}.  
In Figure\,\ref{fig:Sn_Sm} (bottom) we show $S_M$ vs. \Mdyn\ for our sample and \citet{Harris13} compilation. 

Similar to the results for $S_N$, the relic galaxy sample lies below the relation for "normal" mass-matched ETGs in the $S_M$--\Mdyn\ plot. Again, this reinforces the notion that relic galaxies are maximally efficient in forming stars relative to GCs, and the process that might be responsible for lower this efficiency (i.e, satellite accretion) are less important in these galaxies.

\begin{figure*}
\centering
\includegraphics[width=1.\textwidth]{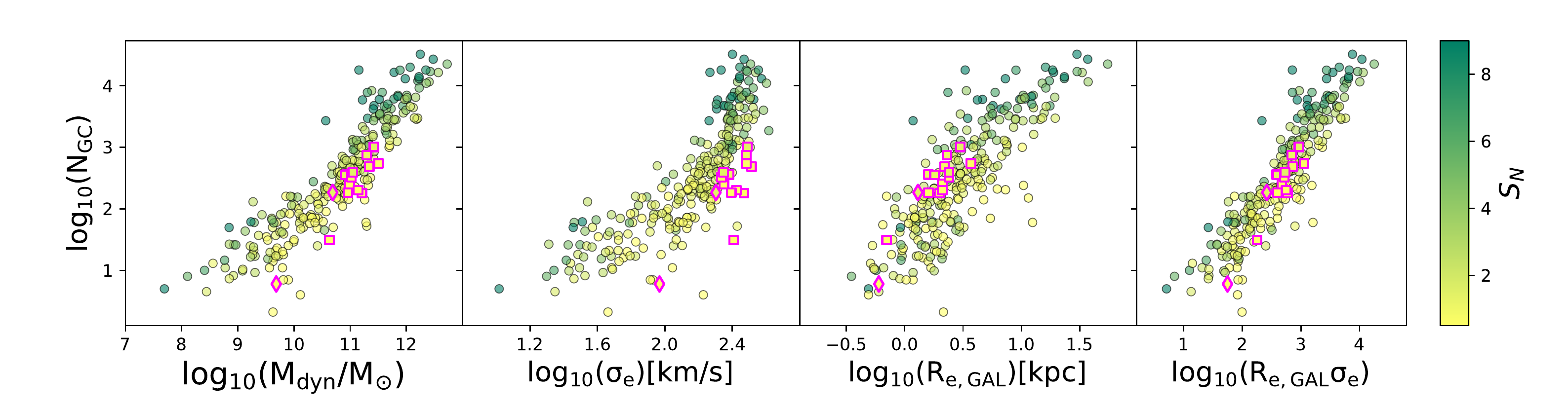}
\caption{Number of GCs versus dynamical mass, velocity dispersion and effective radii of their host galaxies. Circles are from Harris \etal2013, while the squares (relics) and diamonds (non-relics) are the galaxies studied in this work. The colour-code indicates the $S_N$. }
\label{fig:NGC_harris}
\end{figure*}


By exploring the relation between \ngc\ and \Mdyn\ for different morphological types, \citet{Harris13} found that for ellipticals with \M$~>10^{10}$\mo\ the number of GCs increases almost in direct proportion to \Mdyn, while S0s and spirals show an offset of 0.2 and 0.3 dex, respectively, below the trend of ellipticals. They claim that for the same mass, the disky galaxies (higher angular momentum) have a lower fraction of GCs, or have higher fraction of field stars. 
Furthermore, \citet{Harris13} found that (\ReGal$\sigma_{\rm e}$)$^{1.3}$ is an accurate predictor of the total number of GCs over the entire mass range, from dwarfs to giants.

In Figure\,\ref{fig:NGC_harris} we show \ngc\ versus \Mdyn, \ReGal, $\sigma_{\rm e}$, and \ReGal$\sigma_{\rm e}$ for our sample and \citet{Harris13} compilation without applying any offset according the morphological type as their Figure\,9. We can see that indeed, our sample, which is biased towards very high velocity dispersions, is dominated by low $S_N$ values, which is a consequence of either a smaller fraction of GCs, or a higher fraction of field stars. Again, the minimal accretion of low mass, high $S_N$ satellites could result in this higher fraction of field stars.
Nevertheless, it is worthwhile mentioning that the offset \ngc\ versus \Mdyn~ for late-type galaxies vs. ETGs could also be a direct result of using such a virial mass estimator. Late-type galaxies have far more of their orbital energy tied up in rotationally supported orbits, so by using Eq. 6 we are likely to bias their dynamical mass estimates to the low side if we do not correct for this in a more detailed dynamical model.
 
\subsection{GCS effective radius correlations}

\begin{figure*}
\centering
\includegraphics[width=0.8\textwidth]{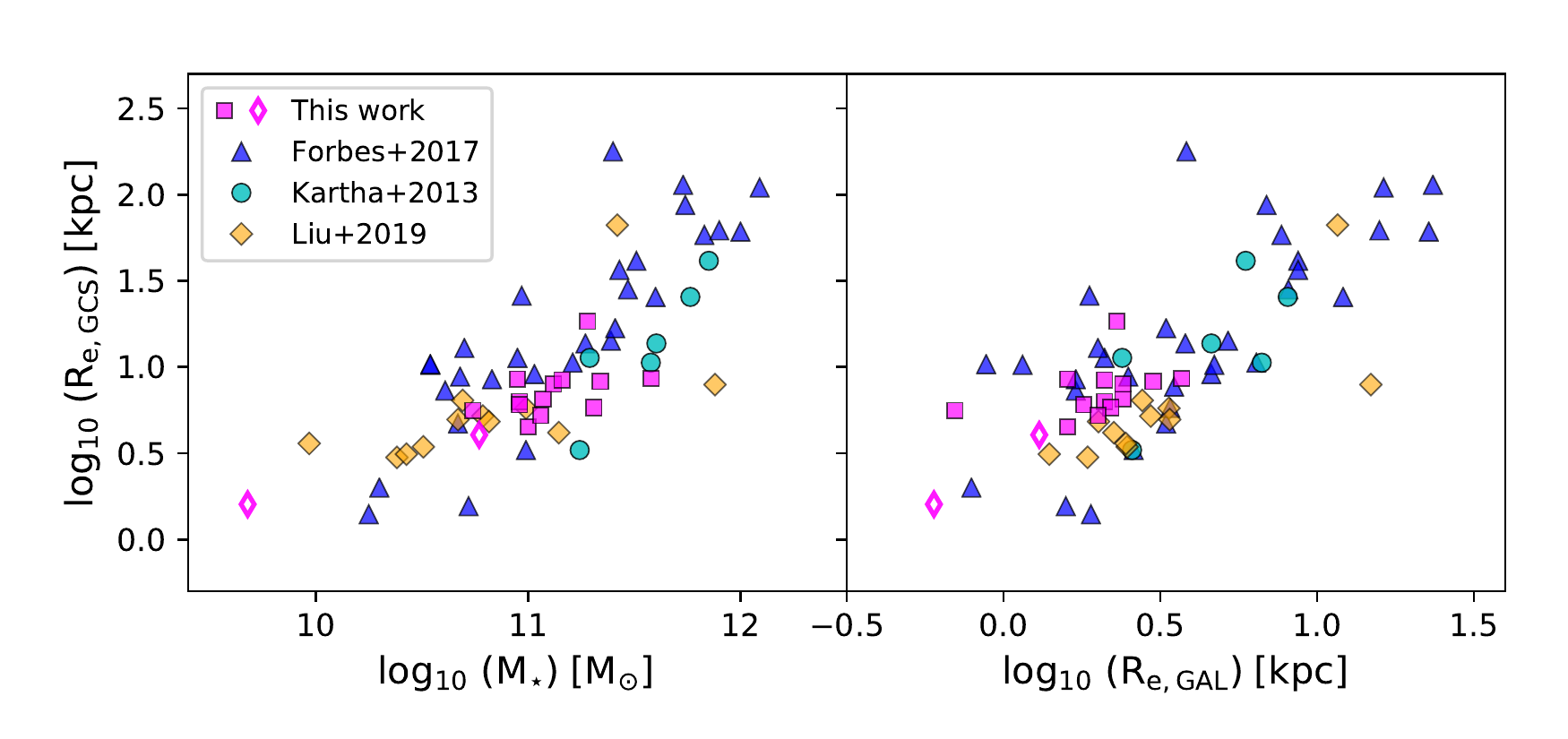}
\caption{GCS effective radius, \Regc\ as function of: \ReGal\ ({\it left}) and stellar mass of the galaxy ({\it right}).}
\label{fig:ReGCS}
\end{figure*}

Numerical simulations predict that accreted stars are preferentially deposited at large galactocentric distances (\citealt*{Naab2009}; \citealt{Font2011}; \citealt{Navarro-Gonzalez2013}). This has been supported observationally from metallicity and age radial gradients (\citealt{Greene2015}; \citealt{Pastorello2014}), as well as kinematics and low surface brightness structures (e.g., \citealt{mackey13}). As the accreted galaxies  not only contain stars but also GCs, the deposition of stars and GCs at large galactocentric distances causes increases in both \Regc\ and \ReGal. 

Although  \Regc\ is commonly measured to correct for spatial incompleteness, it is a GCS property little explored.  
Studies of the spatial extension of the GCS, measured as the radial distance at which the GC surface density reaches the background levels (\citealt{Rhode2007}; \citealt*{Rhode2010}; \citealp{Kartha14}) showed that the extension of the GCS is proportional to the host galaxy stellar mass and \ReGal (but see also \cite{saif20} for the case of ultra-diffuse galaxies). However, as pointed out by \cite{Kartha14}, the extension of the GCS is strongly dependent on the quality of the data and the width of FOV, and claimed that the correlations should be considered more as a general trend than a quantitative relation.
Similarly, studies of \Regc\ scaling relations (\citealt{Forbes2017}; \citealt{Hudson_Robison2018}) reported not unexpectedly correlations between \Regc\ and host galaxy properties. However, although is consistent the claim that \Regc\ scales with \ReGal, the scaling factor is significantly different among authors. 

In Figure\,\ref{fig:ReGCS} we show \Regc\ as function of \ReGal\ and stellar mass of the host galaxy for our sample and data from the literature. 
Our data follow the general trend where the spatial extension of the GCS is proportional to galaxy stellar mass and \ReGal\ but with a closer distribution to ETGs in the Fornax galaxy cluster (\citealt{Liu2019}). 
We note that there is no evidence for stripping in these relic candidates, i.e., all of the correlations between the GC and system sizes reflect the co-evolution of both due to merging (or lack thereof) modulating the orbits/sizes of both the field stars and GCs.


\begin{table*}
    \begin{threeparttable}
	\centering
	\caption{Globular cluster system parameters: (1) Galaxy name. (2) Total number of GCs. (3) The radius that contains half of the GC population. (4) Surface number density of GCs at \Regc. (5) \sersic\ index. (6) Surface number density of background point sources. (7) Specific frequency. (8) Mean $I-H$. (9) Standard deviation of $I-H$. (10) Skewness of $I-H$. (11) Fraction of detected to total GC number (after correction by incompleteness in area and photometry). }
	\label{tab:Table2_GCS}
	\begin{tabular}{lrrrrrrrrrr}
\hline
 Galaxy    &          \ngc &       \Regc[kpc] &    $\Sigma_0$ &       $n_{
m GCS}$ &   $BG_{ps}$ &   $S_N$ &   $\overline{I-H}$\tnote{*} &   $\sigma_{I-H}$\tnote{*} &   skewness\tnote{*,a} &   detected/total \\
(1) & (2) &  (3) & (4) &  (5) &  (6) &  (7) & (8) &   (9) &   (10) &  (11) \\
\hline

NGC 0384  &  356 $\pm$ 47 &   6.04 $\pm$ 8.7 & 0.015 & 1.1 &  0.009 &   1.21 &  0.64 &  0.09 &      - & 0.28 \\
NGC 0472  &  328 $\pm$ 37 &   6.52 $\pm$ 0.8 & 0.021 & 0.5 &  0.003 &   0.74 &  0.59 &  0.11 &     - & 0.23 \\
MRK 1216  & 1013 $\pm$ 80 &   8.24 $\pm$ 0.7 & 0.034 & 0.5 &  0.003 &   1.43 &  0.59 &  0.10 &     -0.66 & 0.11 \\
NGC 1270  &  485 $\pm$ 64 &   5.82 $\pm$ 0.8 & 0.032 & 0.5 &  0.008 &   0.88 &  0.78 &  0.13 &     -0.67 & 0.22 \\
NGC 1271  &  180 $\pm$ 84 &   5.23 $\pm$ 1.1 & 0.018 & 0.7 &  0.003 &   0.27 &  0.69 &  0.09 &      - & 0.22 \\
NGC 1281  &  360 $\pm$ 30 &    4.5 $\pm$ 0.5 & 0.035 & 0.7 &  0.004 &   1.13 &  0.70 &  0.09 &      - & 0.27\\
NGC 1282  &  184 $\pm$ 27 &   4.04 $\pm$ 1.5 & 0.017 & 0.6 &  0.005 &   1.37 &  0.58 &  0.11 &      - & 0.82 \\
UGC 2698  & 547 $\pm$ 100 &   8.57 $\pm$ 1.4 & 0.026 & 0.5 &  0.004 &   0.62 &  0.64 &  0.14 &     -0.59 & 0.18 \\
NGC 2767  &  389 $\pm$ 49 &   7.96 $\pm$ 2.5 & 0.017 & 0.8 &  0.002 &   1.47 &  0.49 &  0.11 &     - & 0.27 \\ 
UGC 3816  &  253 $\pm$ 20 &   6.31 $\pm$ 2.5 & 0.017 & 0.8 &  0.004 &   0.55 &  0.51 &  0.09 &      0.47 & 0.49 \\
NGC 3990  &     6 $\pm$ 2 &    1.6 $\pm$ 1.4 & 0.002 & 0.5 &  0.001 &   0.17 &  0.54 &  0.09 &      - & 1.00 \\
PGC 11179 &  200 $\pm$ 43 &   8.42 $\pm$ 3.0 & 0.005 & 0.5 &  0.003 &   0.42 &  0.65 &  0.10 &     -0.70 & 0.09 \\
PGC 12562 &   31 $\pm$ 13 &    5.6 $\pm$ 8.8 & 0.004 & 0.5 &  0.007 &   0.13 &  0.84 &  0.21 &      - & 0.44 \\
PGC 32873 & 745 $\pm$ 238 & 18.4 $\pm$ 119.5 & 0.004 & 1.5 &  0.000 &   1.06 &  0.42 &  0.07 &     - & 0.06 \\
PGC 70520 &  183 $\pm$ 70 &   8.52 $\pm$ 2.4 & 0.007 & 0.5 &  0.003 &   0.36 &  0.51 &  0.08 &      - & 0.26 \\
\hline
\end{tabular}
\begin{tablenotes}
   \item[*]  Measurments using all the GC candidate sources from 0 to 2\Regc, withouth background contamination subtraction.
   \item[a] Values significantly different from zero at 95\% confidence level. Empty spaces are consistent with zero skew values.
  \end{tablenotes}
	\end{threeparttable}
\end{table*}


\section{Conclusions}

In this study we analyze the GC systems of a sample of 15 massive compact early-type galaxies from which 13 are relic galaxy candidates (\citealt{y17}). By using archival HST imaging in the $I$ and $H$ bands, we determine the GCLF, \ngc, color and spatial distribution of the GCS, $S_N$, $S_M$ and, look for correlations with host galaxy properties from the literature, such as, \Mstar, \Mdyn, \lamR, stellar velocity dispersion and metallicity. Our main findings are:

\begin{enumerate}
 
    \item The compact galaxy sample has low GC specific frequencies, $S_N<2.5$ with a median of $S_N=1$, whereas normal ETGs of the same mass typically have $2<S_N<10$.
    This is in agreement with the picture that the galaxies in our sample experienced high star-formation efficiencies as would be the case for a rapid, early dissipative formation, together with relatively low levels of accretion of high-$S_N$, low-mass satellites.    
    
    \item The GCS spatial distributions  are similar  to those in normal ETGs in that they are more extended than the starlight,  and \Regc\ correlates with the galaxy properties \ReGal, \Mstar\ and \Mhalo. 

    
 
     
     \item Intriguingly, we find a mild, but significant anti-correlation between the standard deviation of the $I-H$ colour distribution and the galaxy specific angular momentum, \lamR. While the present dataset is relatively small, if confirmed this result might be expected from hierarchical merging models whereby galaxies which have undergone less accretion/merging activity might be expected to preserve their initial \lamR\ and have less complex GCS colour distributions.
     
\end{enumerate}

Future simulations can help constrain the diagnostic power of \lamR~ when allied to the width of the colour distributions of GC systems for  understanding of  merger histories of galaxies. Moreover,
quantification of the globular cluster color distributions of this sample with optical HST photometry would be extremely valuable to better constrain the accreted mass fractions in these systems. 

\section*{Acknowledgements}
KAM acknowledges funding from CAPES/PNPD and CONACyT. 
ACS acknowledges funding from the brazilian agencies \textit {Conselho Nacional de Desenvolvimento Cient\'ifico e Tecnol\'ogico} (CNPq) and the Rio Grande do Sul Research Foundation (FAPERGS) through grants CNPq-403580/2016-1, CNPq-11153/2018-6, PqG/FAPERGS-17/2551-0001, FAPERGS/CAPES 19/2551-0000696-9 and L'Or\'eal UNESCO ABC \emph{Para Mulheres na Ci\^encia}.
MAB acknowledges support from grant AYA2016-77237-C3-1-P from the Spanish Ministry of Economy and Competitiveness (MINECO) and from the Severo Ochoa Excellence scheme (SEV-2015-0548). CF acknowledges the financial support from CNPQ (processes 433615/2018-4 and 311032/2017-6), MT thanks the support of CNPq-307675/2018-1 and 
the program L'Or\'eal UNESCO ABC \emph{Para Mulheres na Ci\^encia}. ASM acknowledges CNPq-308306/2018-0. We acknowledge discussions with Schoennell, Fabrício Ferrari and Rogério Riffel. We acknowledge the reception of a fruitful week in the peaceful countryside of Pueblo Mariana.
This research has made use of the NASA/IPAC Extragalactic Database, which is funded by the National Aeronautics and Space Administration and operated by the California Institute of Technology.

\section*{Data Availability}

The data underlying this article are available from the corresponding author, upon reasonable request.




\bibliographystyle{mnras}
\bibliography{./references_KA} 


\appendix

\section{Individual GC systems}
\begin{figure*}
\centering
\includegraphics[width=0.8\textwidth]{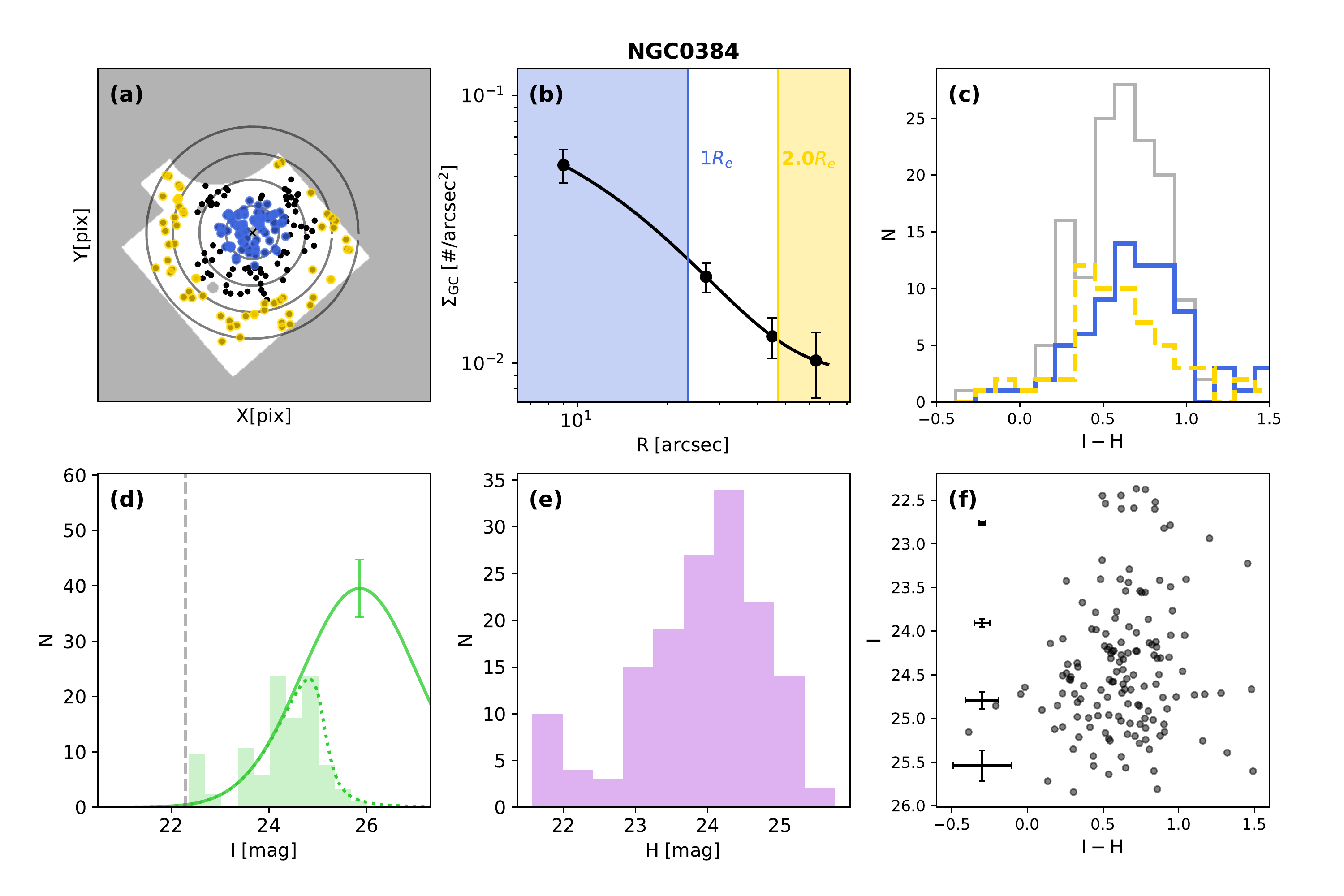}
\caption{Derived parameters for the individual GC system of NGC\,384 (a) the spatial distribution of the GC candidates highlighting with blue and yellow the sources belonging to the {\it inner} and {\it outer} population  according to the \sersic\ profile (panel b). The gray regions show masked areas due to bright stars, galaxies and pixels with bad quality. The circles show the annuli on which the surface number density of GCs was determined. 
(b) Surface number density of the GC candidates as function of radius. The solid line shows the best \sersic\ fit. 
(c) $I-H$ color distributions for the population within 1\Regc\ (solid blue), outer to 2\Regc\ (dashed yellow), and within 2\Regc\ (solid gray). 
(d) I-band luminosity function of the {\it inner} population corrected for contaminants. The dotted line shows the best Gaussian*Pritchet function, the solid line is the corresponding single Gaussian. 
(e) H-band luminosity function of the GC candidates within 2${R_e}$. 
(f) $I-H$ vs. $I$ color magnitude diagram of the GC candidates within 2${R_e}$ with the mean error bars per magnitude bin.}
\label{fig:A_NGC0384}
\end{figure*}

\begin{figure*}
\centering
\includegraphics[width=0.8\textwidth]{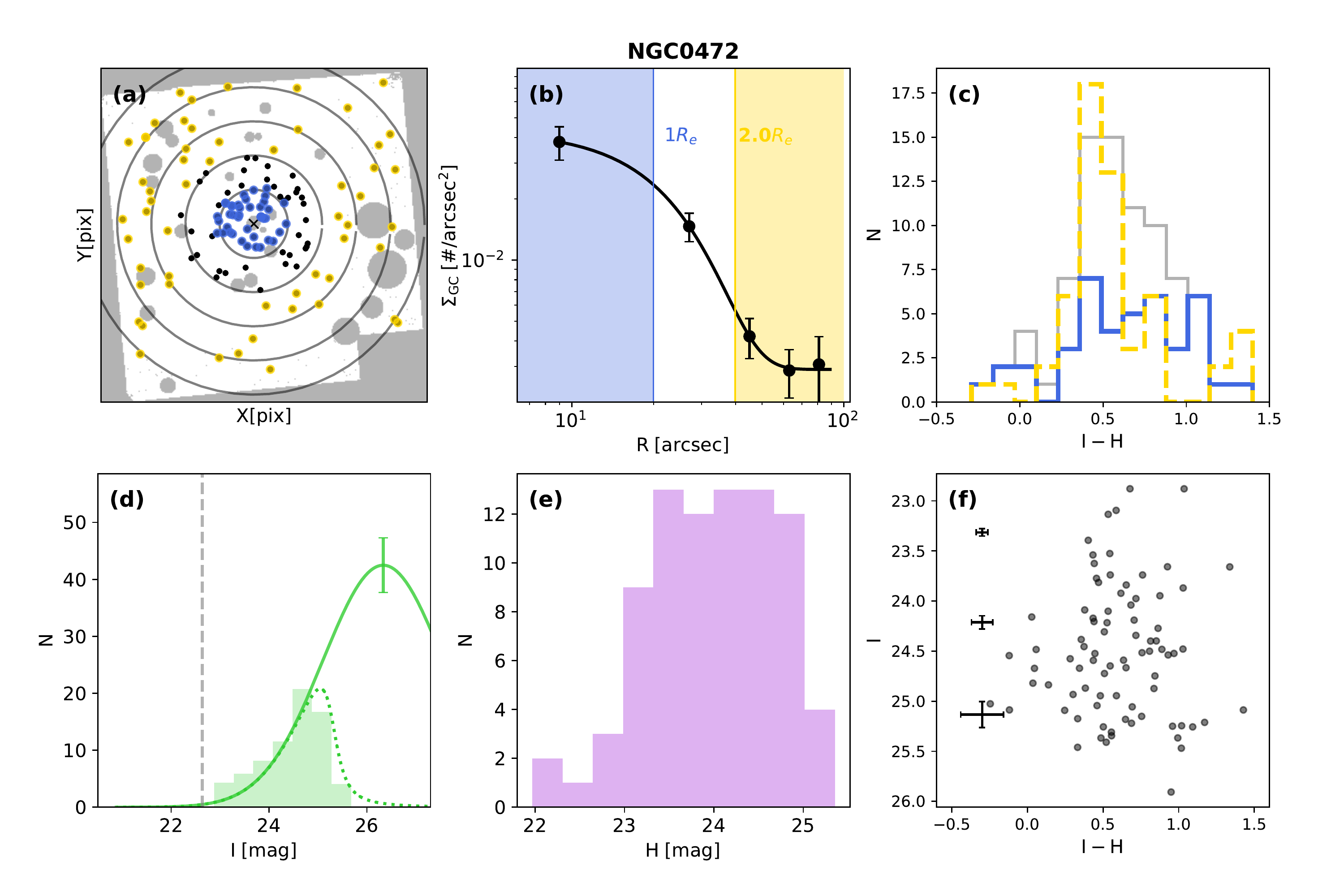}
\caption{Same as in Fig~\ref{fig:A_NGC0384}.}
\label{fig:A_NGC0472}
\end{figure*}

\begin{figure*}
\centering
\includegraphics[width=0.8\textwidth]{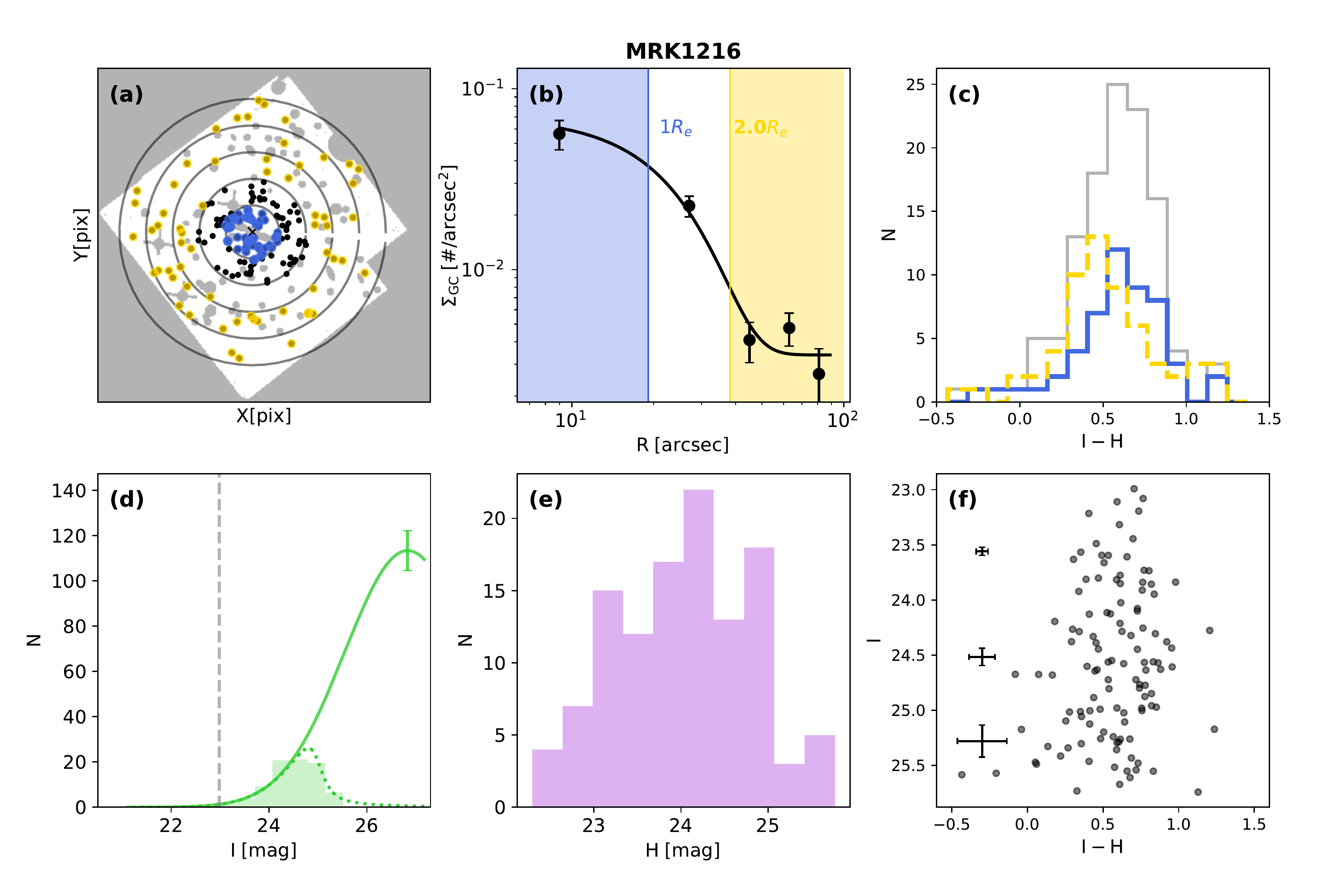}
\caption{Same as in Fig~\ref{fig:A_NGC0384}.}
\label{fig:A_MRK1216}
\end{figure*}

\begin{figure*}
\centering
\includegraphics[width=0.8\textwidth]{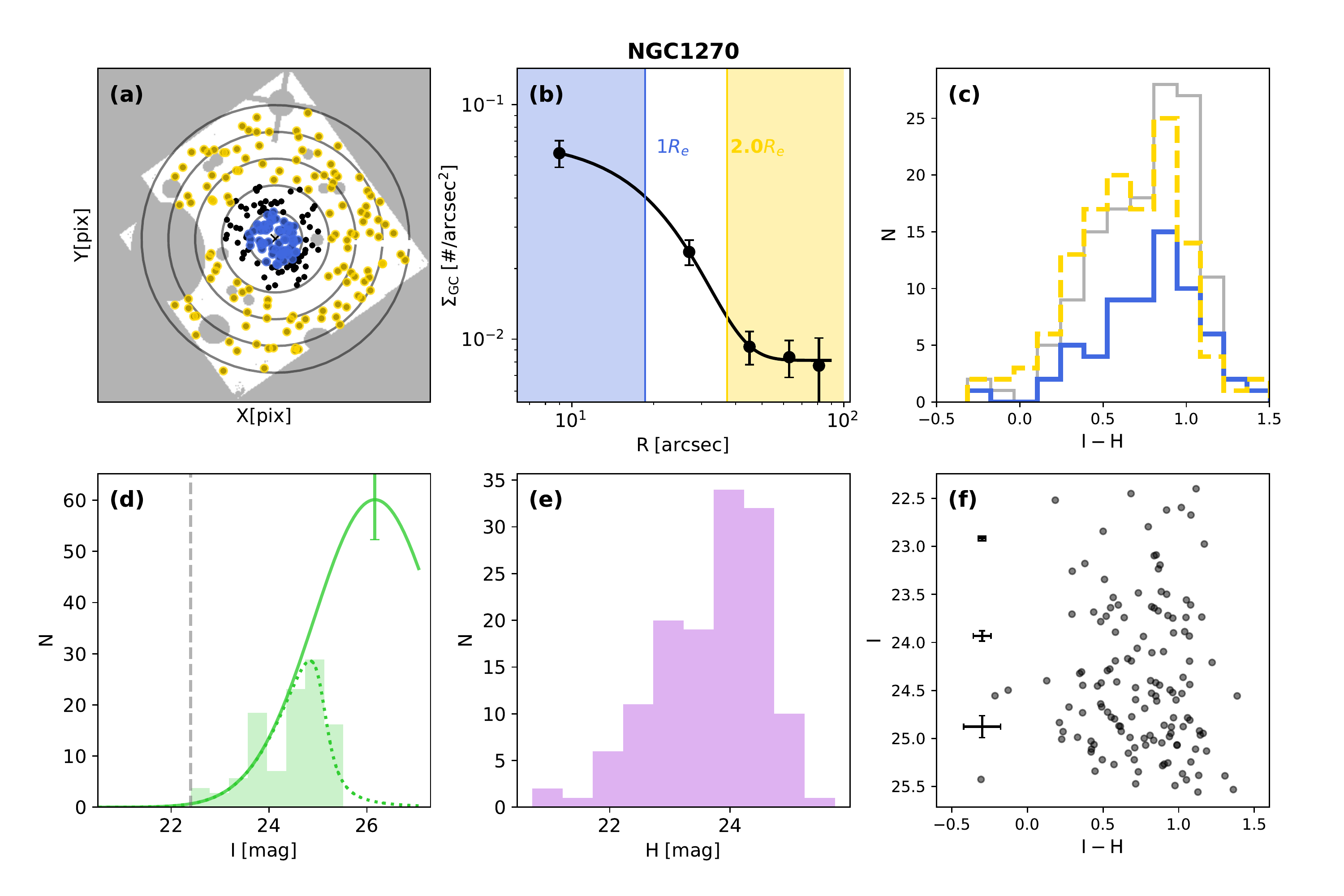}
\caption{Same as in Fig~\ref{fig:A_NGC0384}.}
\label{fig:A_NGC1270}
\end{figure*}

\begin{figure*}
\centering
\includegraphics[width=0.8\textwidth]{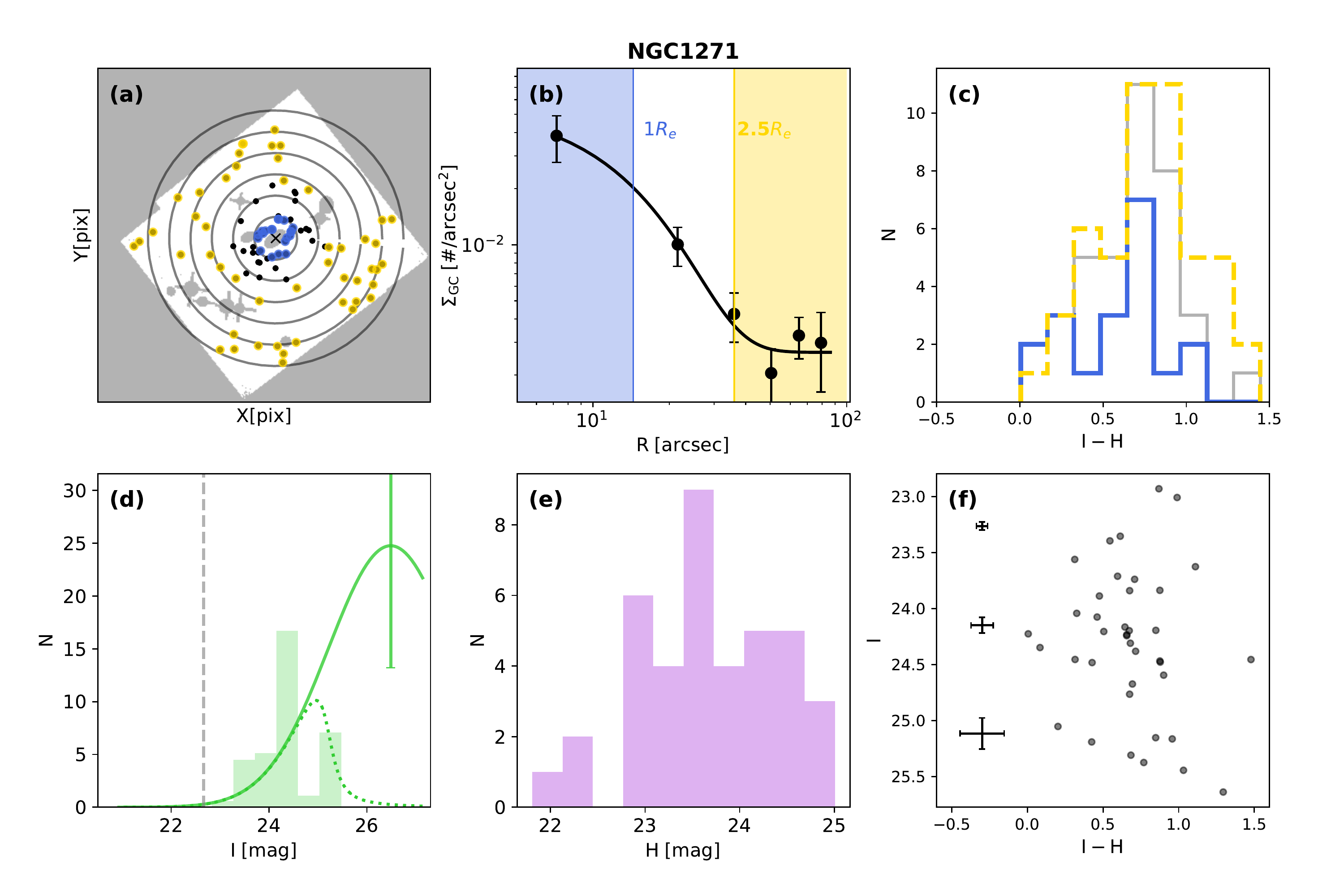}
\caption{Same as in Fig~\ref{fig:A_NGC0384}.}
\label{fig:A_NGC1271}
\end{figure*}

\begin{figure*}
\centering
\includegraphics[width=0.8\textwidth]{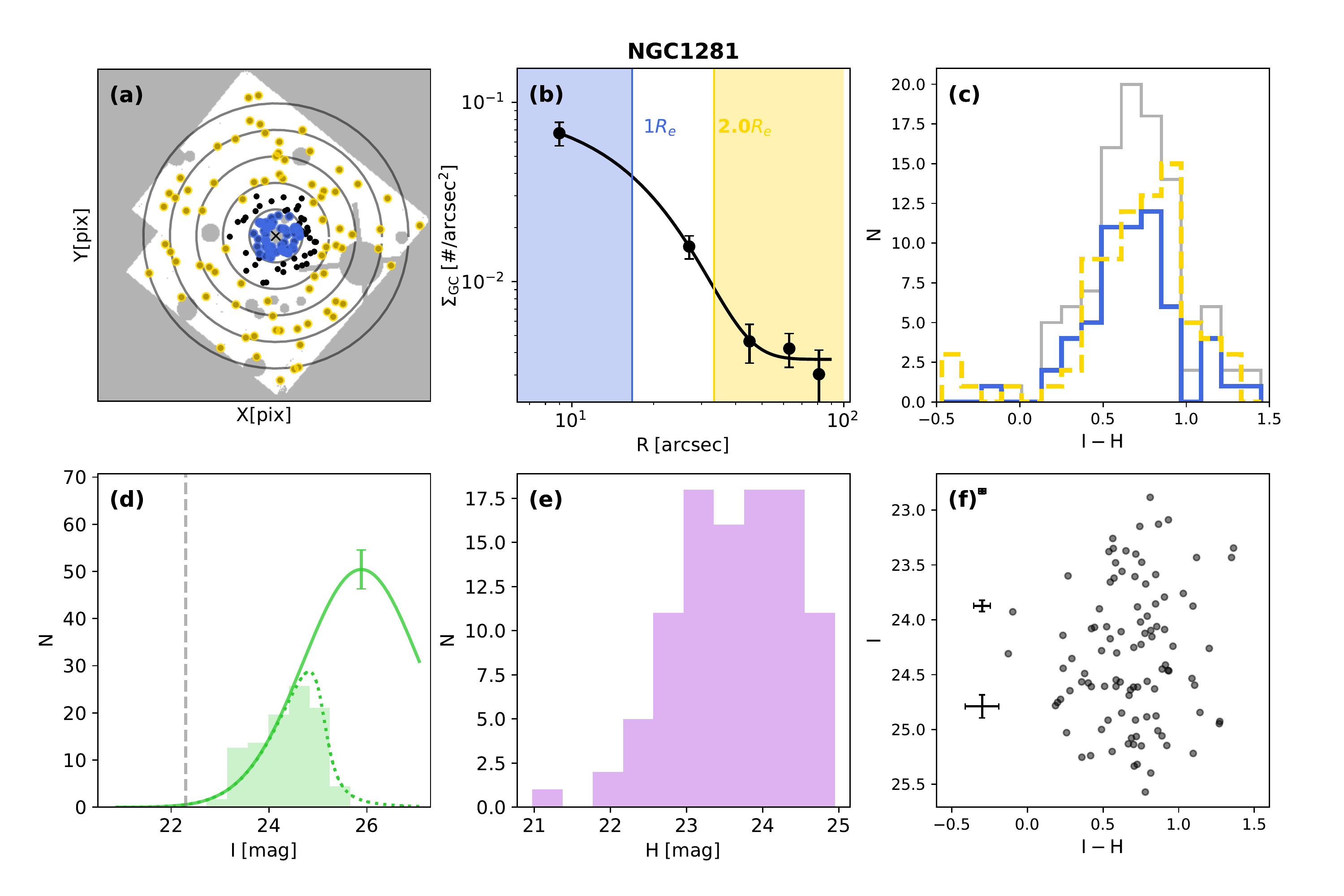}
\caption{Same as in Fig~\ref{fig:A_NGC0384}.}
\label{fig:A_NGC1281}
\end{figure*}

\begin{figure*}
\centering
\includegraphics[width=0.8\textwidth]{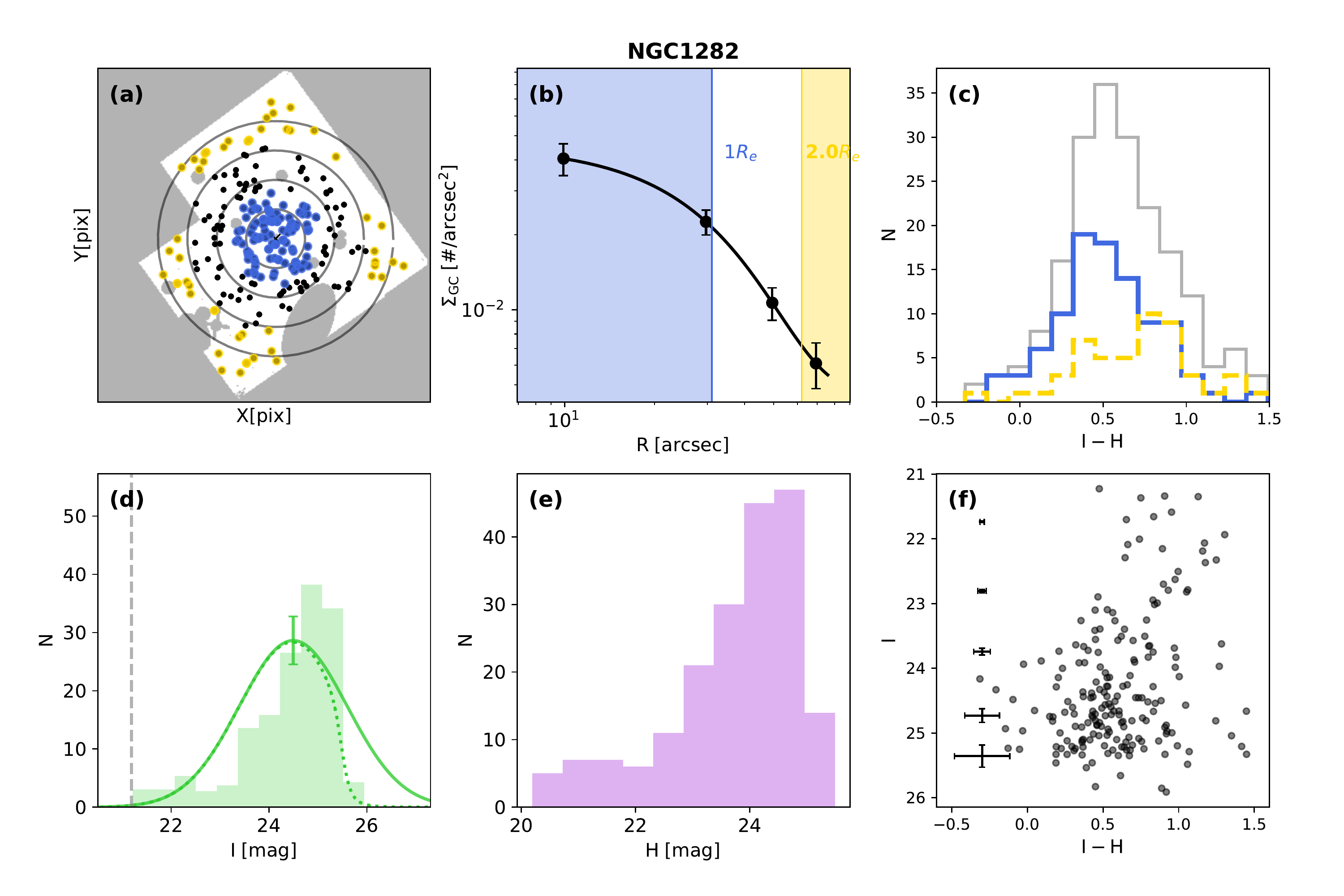}
\caption{Same as in Fig~\ref{fig:A_NGC0384}.}
\label{fig:A_NGC1282}
\end{figure*}

\begin{figure*}
\centering
\includegraphics[width=0.8\textwidth]{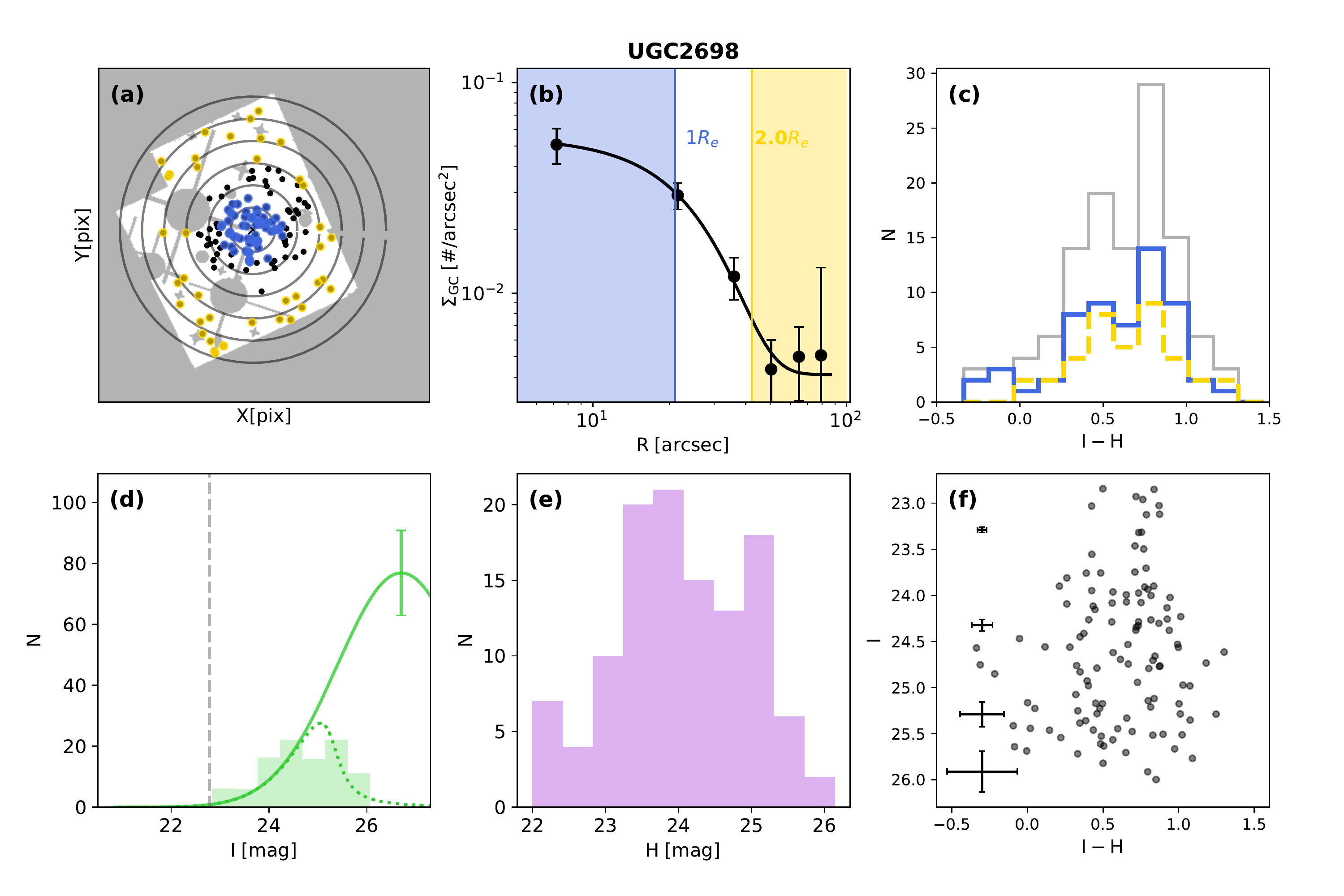}
\caption{Same as in Fig~\ref{fig:A_NGC0384}.}
\label{fig:A_UGC2698}
\end{figure*}

\begin{figure*}
\centering
\includegraphics[width=0.8\textwidth]{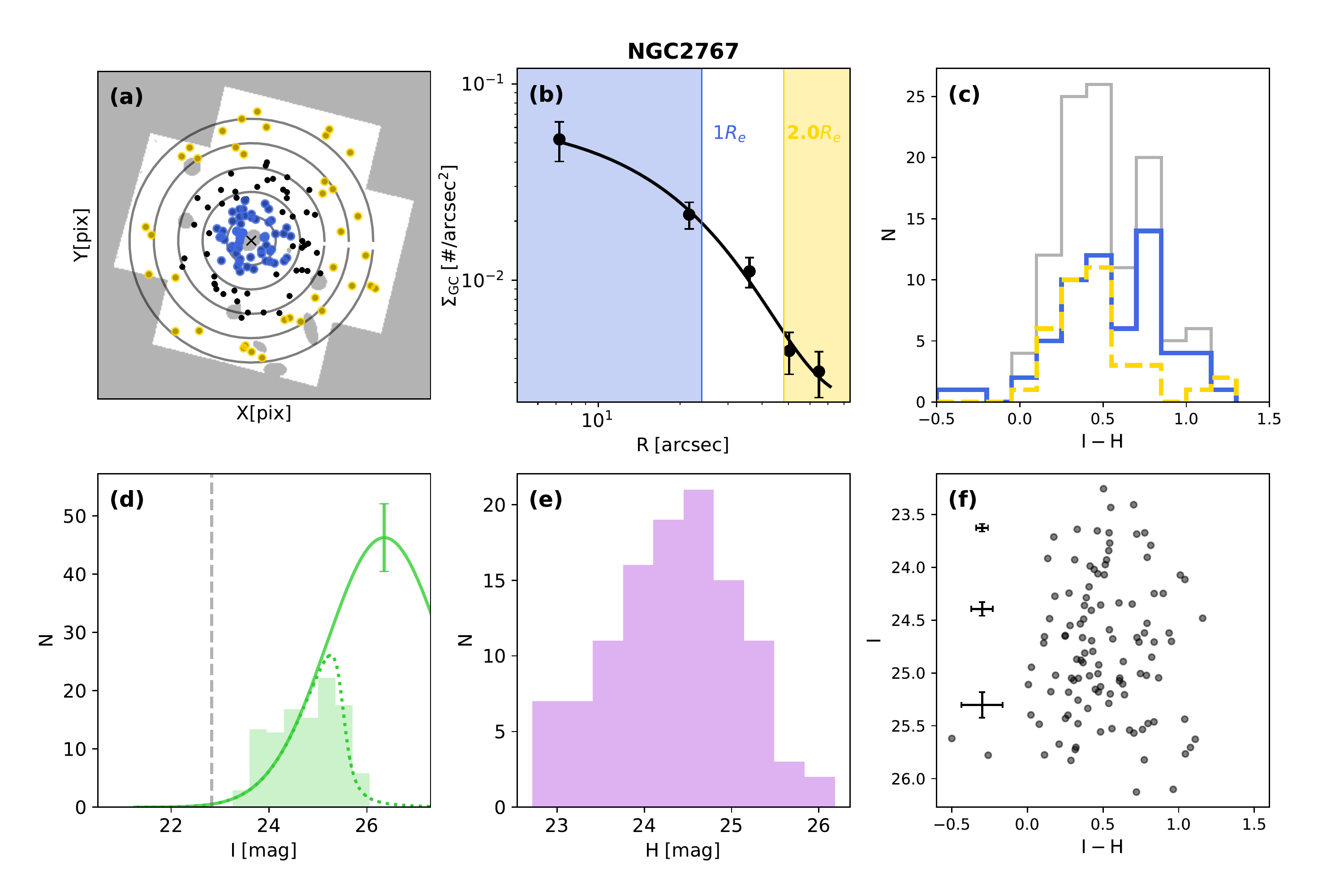}
\caption{Same as in Fig~\ref{fig:A_NGC0384}.}
\label{fig:A_NGC2768}
\end{figure*}

\begin{figure*}
\centering
\includegraphics[width=0.8\textwidth]{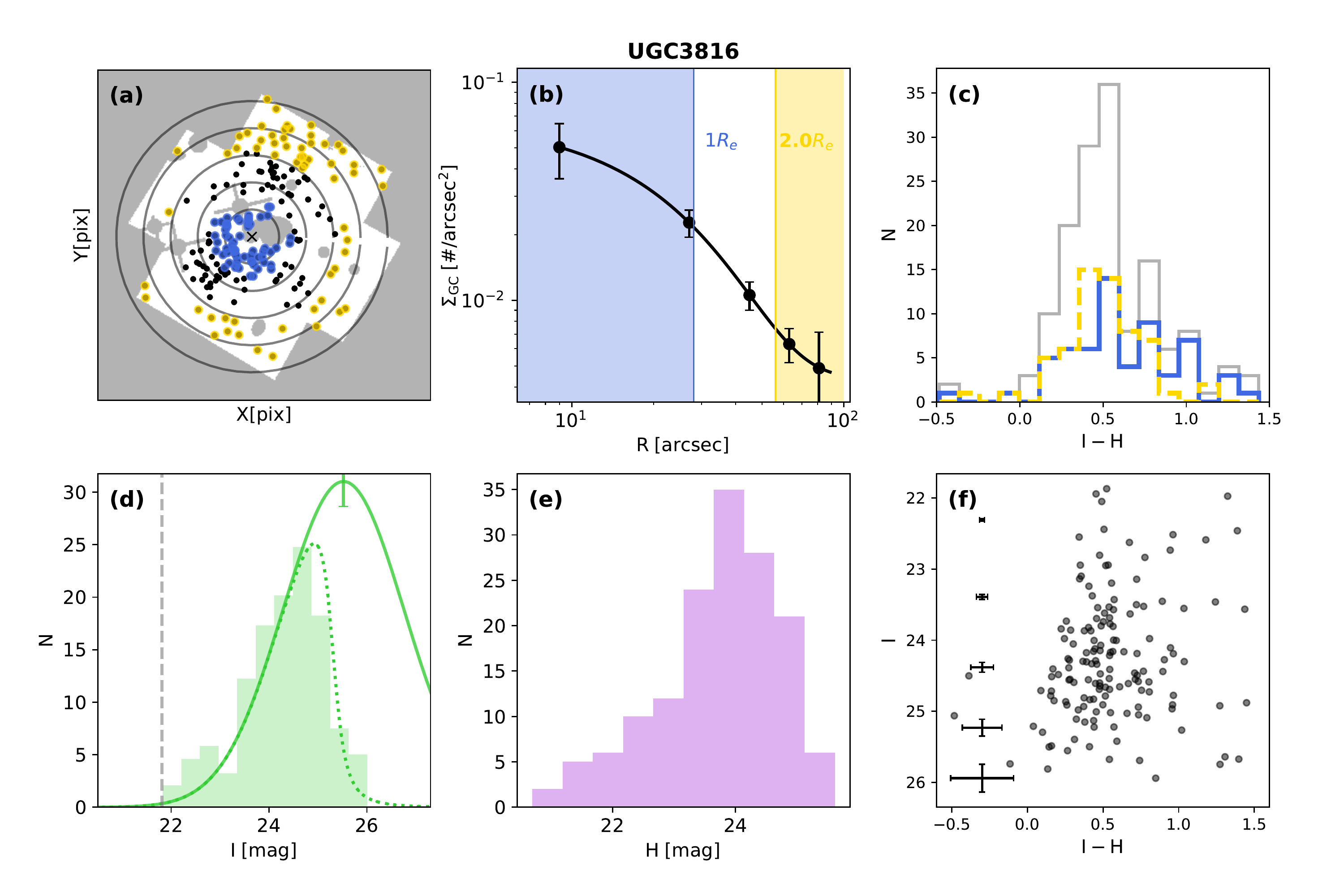}
\caption{Same as in Fig~\ref{fig:A_NGC0384}.}
\label{fig:A_UGC282}
\end{figure*}

\begin{figure*}
\centering
\includegraphics[width=0.9\textwidth]{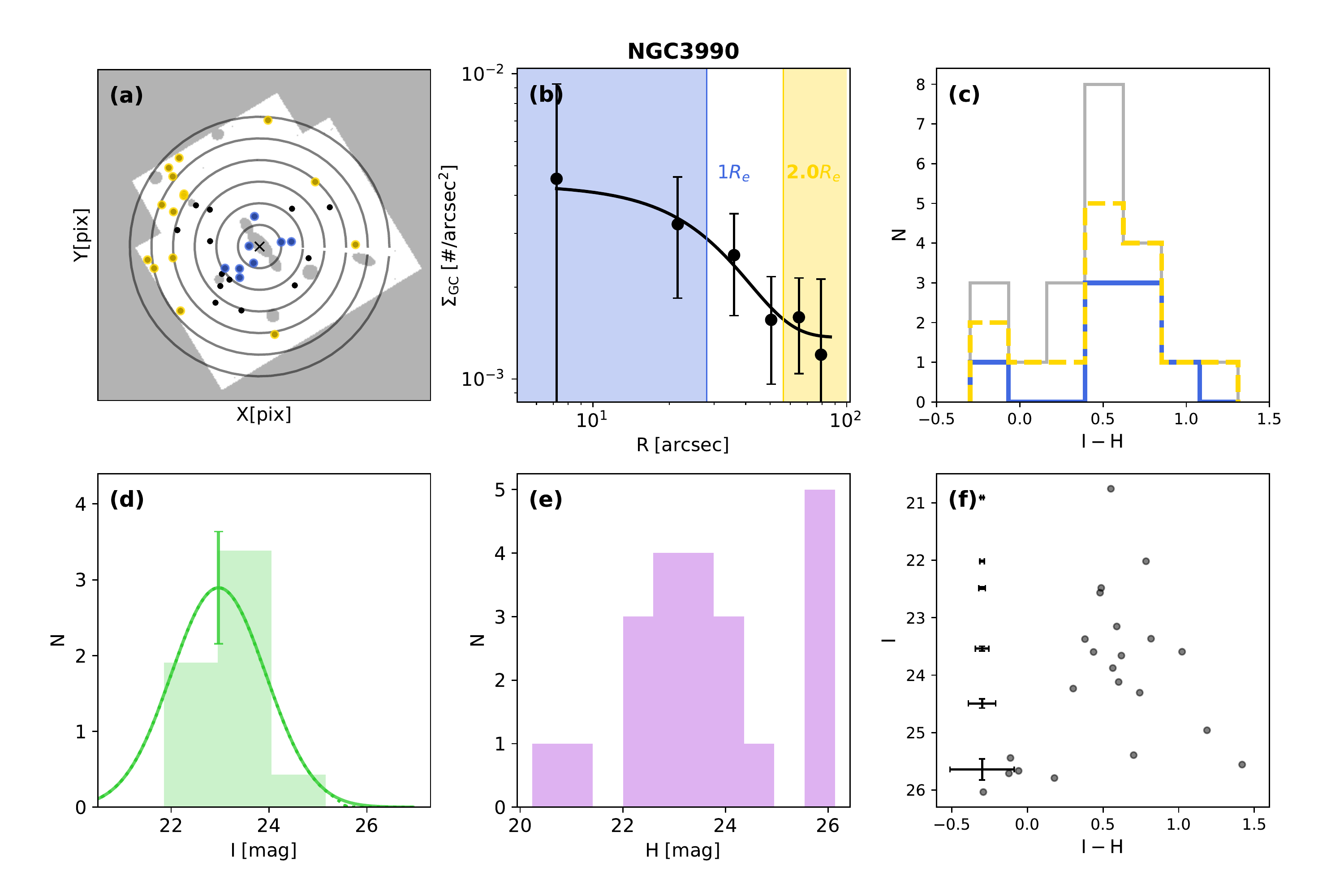}
\caption{Same as in Fig~\ref{fig:A_NGC0384}.}
\label{fig:A_NGC3990}
\end{figure*}

\begin{figure*}
\centering
\includegraphics[width=0.8\textwidth]{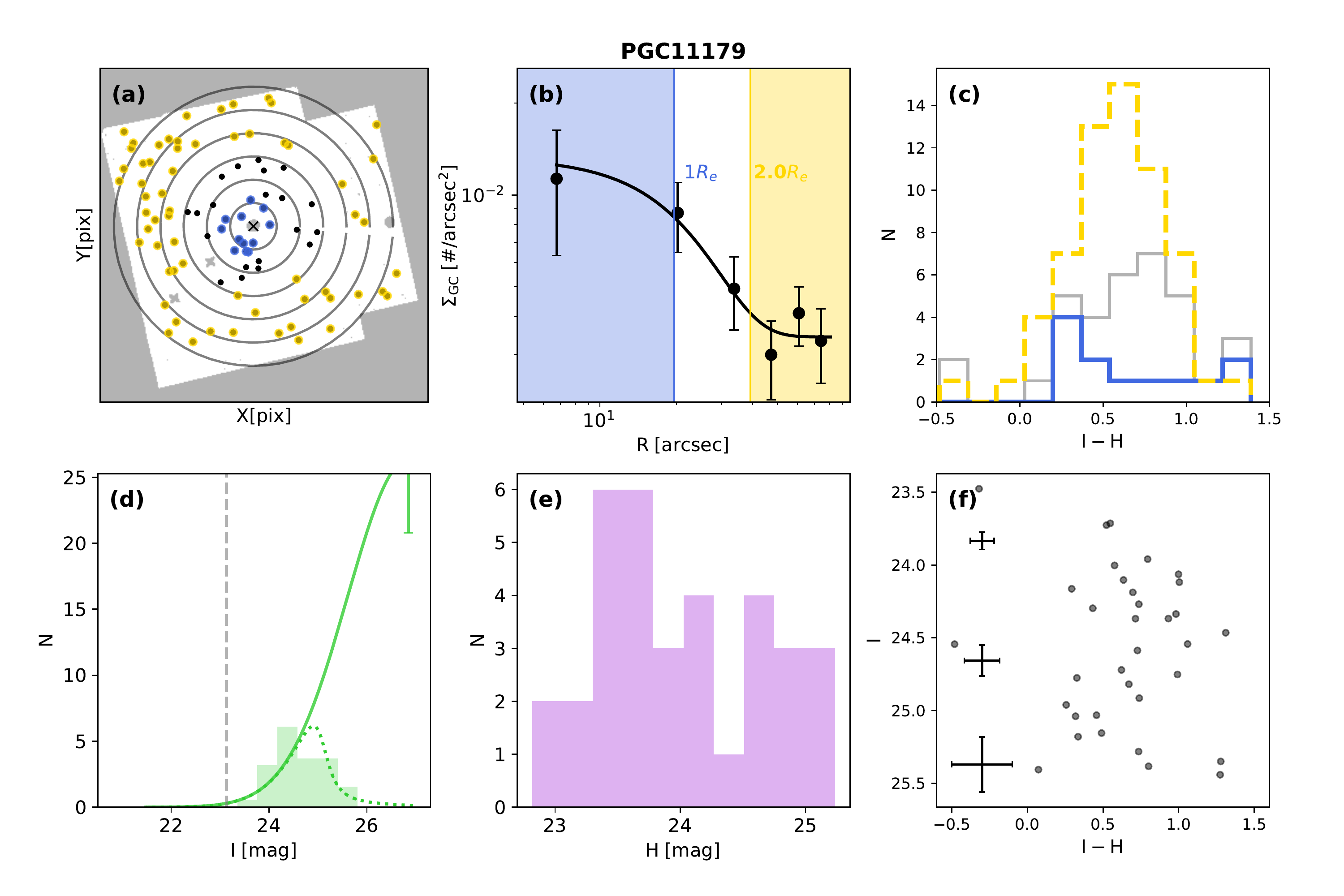}
\caption{Same as in Fig~\ref{fig:A_NGC0384}.}
\label{fig:A_PGC11179}
\end{figure*}

\begin{figure*}
\centering
\includegraphics[width=0.8\textwidth]{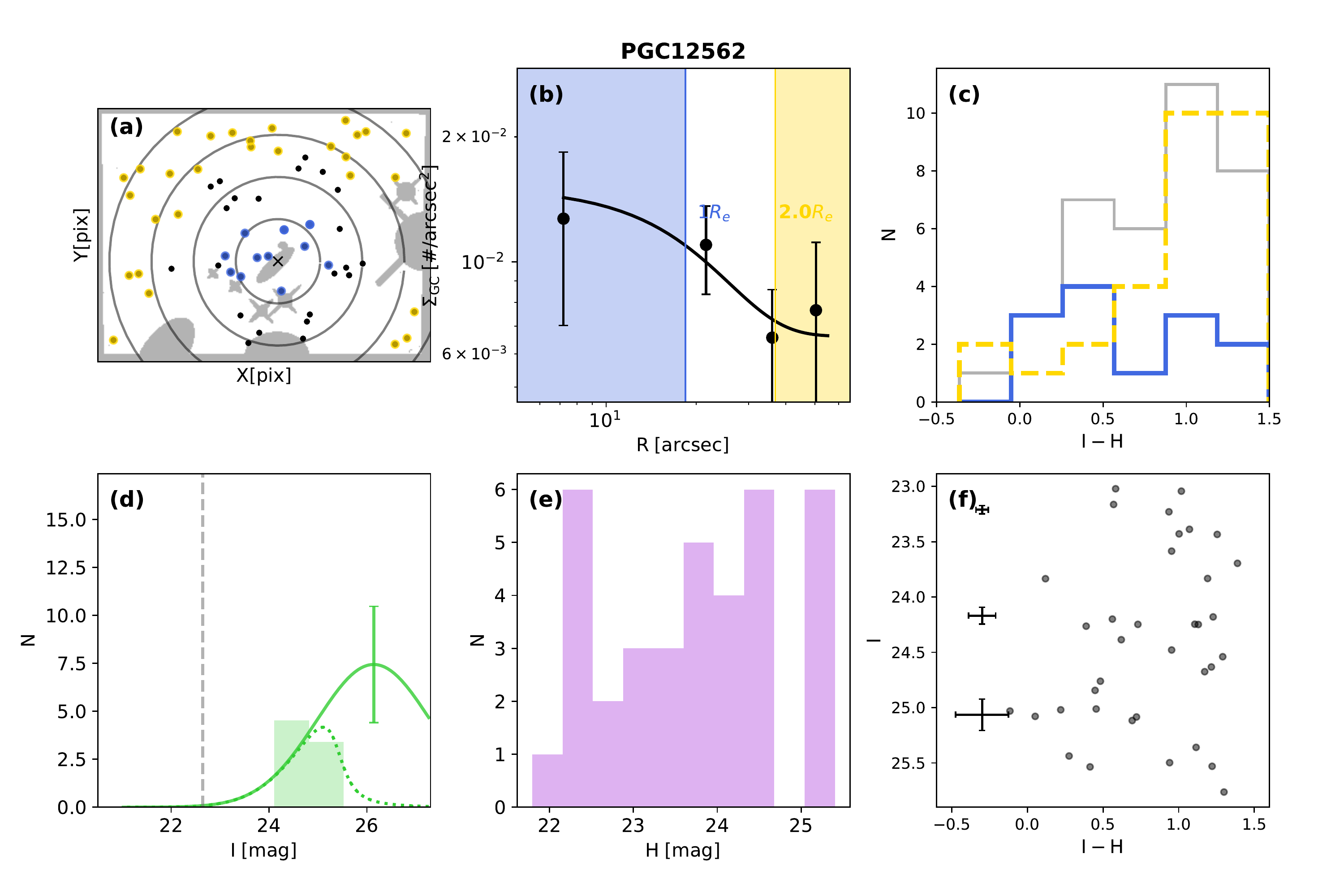}
\caption{Same as in Fig~\ref{fig:A_NGC0384}.}
\label{fig:A_PGC12562}
\end{figure*}

\begin{figure*}
\centering
\includegraphics[width=0.8\textwidth]{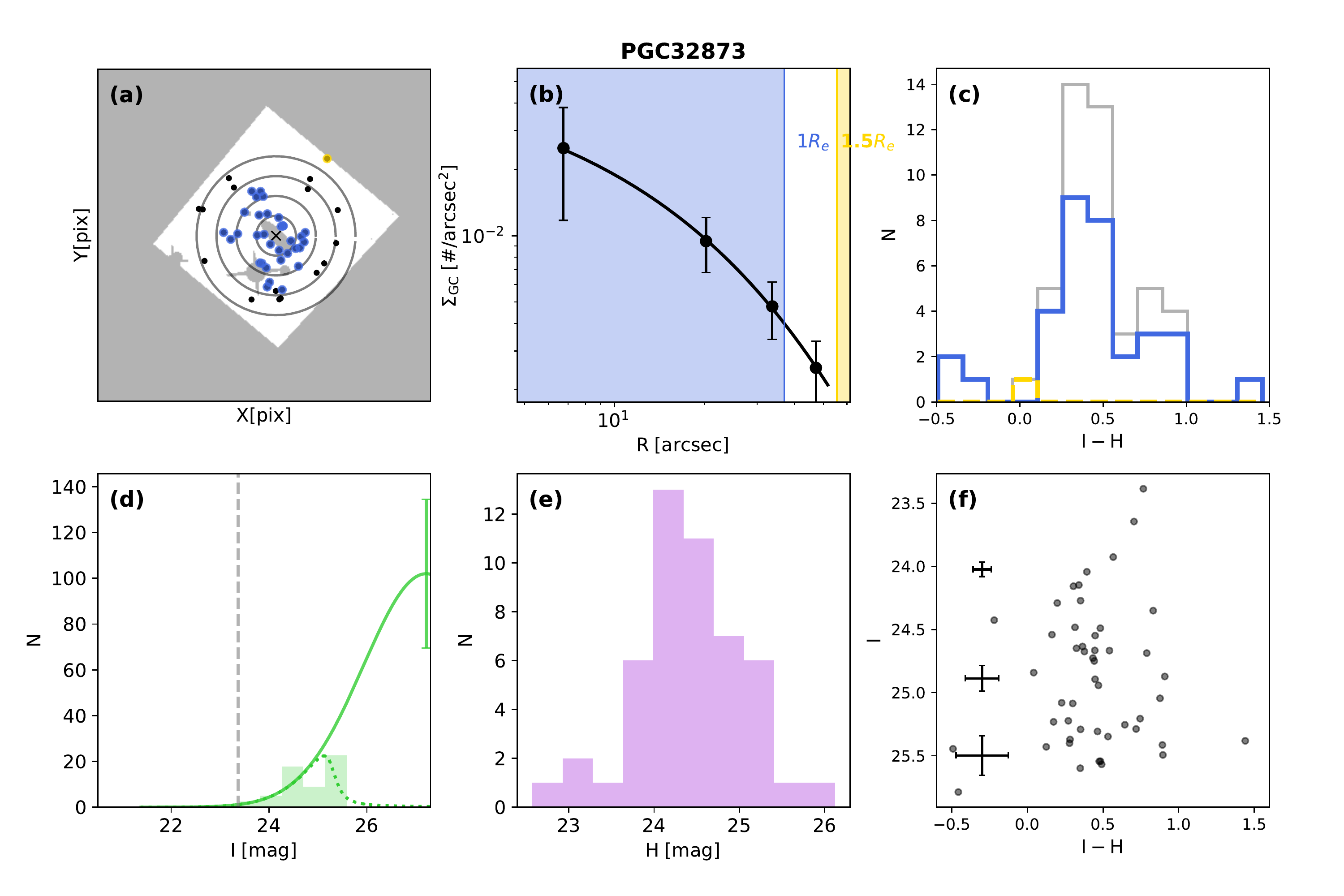}
\caption{Same as in Fig~\ref{fig:A_NGC0384}.}
\label{fig:A_PGC32873}
\end{figure*}

\begin{figure*}
\centering
\includegraphics[width=0.8\textwidth]{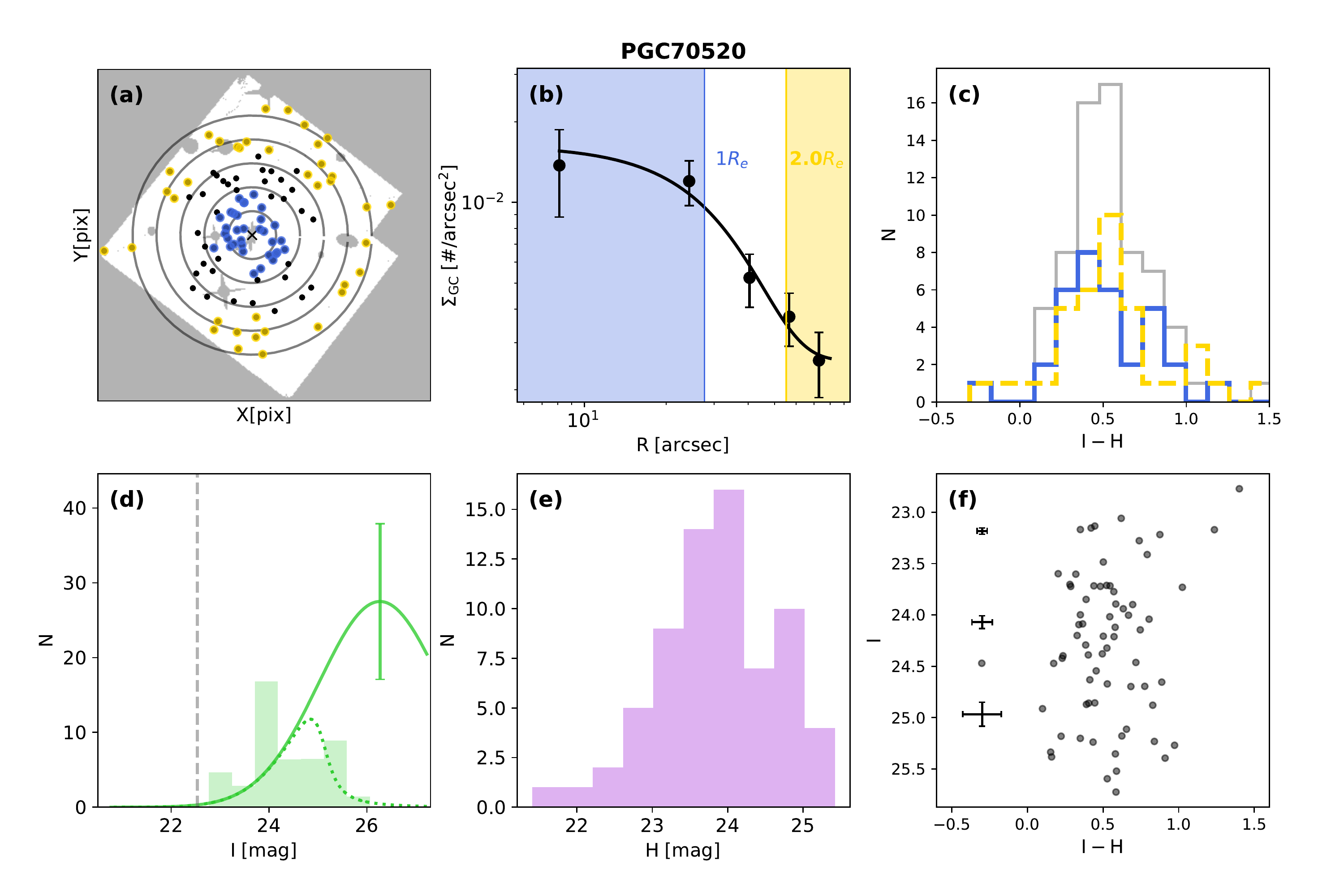}
\caption{Same as in Fig~\ref{fig:A_NGC0384}.}
\label{fig:A_PGC70520}
\end{figure*}

In Figures \ref{fig:A_NGC0384} to \ref{fig:A_PGC70520} we  show for each individual GC system several derived properties: the spatial distribution of the GC candidates, the surface number density of the GC candidates as function of radius with the best \sersic\ fit, 
the $I-H$ color distributions for the population at different galactocentric radii, the I-band luminosity function, the H-band luminosity function of the GC candidates within 2${R_e}$ and the $I-H$ color magnitude diagram of the GC candidates within 2${R_e}$.

\bsp	
\label{lastpage}
\end{document}